\documentclass{aa}
\usepackage{natbib}
\usepackage{graphicx}
\usepackage{ulem}

\usepackage{txfonts}

\usepackage{xcolor}

\newcommand{\code}[1]{\texttt{#1}}

\def\nifs{\iso{56}Ni}
\def\cofs{\iso{56}Co}
\def\fefs{\iso{56}Fe}

\def\cm3{cm$^{-3}$}
\def\kms{\mbox{km~s$^{-1}$}}

\def\msun{$M_{\odot}$}

\def\one{\ts {\,\sc i}}
\def\two{\ts {\,\sc ii}}
\def\three{\ts {\,\sc iii}}
\def\four{\ts {\,\sc iv}}

\def\beq{\begin{equation}}
\def\eeq{\end{equation}}

\def\lesssim{\mathrel{\hbox{\rlap{\hbox{\lower4pt\hbox{$\sim$}}}\hbox{$<$}}}}
\def\gtrsim{\mathrel{\hbox{\rlap{\hbox{\lower4pt\hbox{$\sim$}}}\hbox{$>$}}}}

\def\one{{\,\sc i}}
\def\two{{\,\sc ii}}
\def\three{{\,\sc iii}}
\def\four{{\,\sc iv}}

\def\v1d{{\code{V1D}}}
\def\kepler{{\code{KEPLER}}}
\def\mesa{{\code{MESA}}}
\def\cmfgen{{\code{CMFGEN}}}

\def\mum{\hbox{$\mu$m}}

\def\oidoub{[O\one]\,$\lambda\lambda$\,$6300,\,6364$}
\def\caiidoub{[Ca\two]\,$\lambda\lambda$\,$7291,\,7323$}

\newcommand{\iso}[2]{\ensuremath{^{#1}\rm{#2}}}

\def\aj{AJ}
\def\pasp{PASP}

\def\apj{ApJ}
\def\apjs{ApJS}
\def\apjl{ApJL}
\def\aap{A\&A}
\def\araa{ARA\&A}

\def\mnras{MNRAS}

\def\physrep{Phys.~Rep.}

\def\solphys{Sol.~Phys.}

\begin{document}

 \title{Radiative-transfer modeling of nebular-phase type II supernova}
 \subtitle{Dependencies on progenitor and explosion properties}

 \titlerunning{Modeling of nebular phase spectra for type II SNe}

\author{Luc Dessart\inst{\ref{inst1}}
  \and
   D. John Hillier\inst{\ref{inst2}}
  }

\institute{
Institut d'Astrophysique de Paris, CNRS-Sorbonne Universit\'e, 98 bis boulevard Arago, F-75014 Paris, France.\label{inst1}
\and
    Department of Physics and Astronomy \& Pittsburgh Particle Physics,
    Astrophysics, and Cosmology Center (PITT PACC),  \hfill \\ University of Pittsburgh,
    3941 O'Hara Street, Pittsburgh, PA 15260, USA.\label{inst2}
  }

   \date{Received; accepted}

  \abstract{
  Nebular phase spectra of core-collapse supernovae (SNe) provide critical and unique information on the progenitor massive star and its explosion. We present a set of 1-D steady-state non-local thermodynamic equilibrium radiative transfer calculations of type II SNe at 300\,d after explosion. Guided by results for a large set of stellar evolution simulations, we craft ejecta models for type II SNe from the explosion of a 12, 15, 20, and 25\,\msun\ star. The ejecta density structure and kinetic energy, the \nifs\ mass, and the level of chemical mixing are parametrized. Our model spectra are sensitive to the adopted line Doppler width, a phenomenon we associate with the overlap of Fe\two\ and O\one\ lines with Ly\,$\alpha$ and Ly\,$\beta$. Our spectra show a strong sensitivity to \nifs\ mixing since it determines where decay power is absorbed. Even at 300\,d after explosion, the H-rich layers reprocess the radiation from the inner metal rich layers. In a given progenitor model, variations in \nifs\ mass and distribution impact the ejecta ionization, which can modulate the strength of all lines. Such ionization shifts can quench Ca\two\ line emission. In our set of models, the \oidoub\ doublet strength is the most robust signature of progenitor mass. However, we emphasize that convective shell merging in the progenitor massive star interior can pollute the O-rich shell with Ca, which will weaken the O\one\ doublet flux in the resulting nebular SN II spectrum. This process may occur in Nature, with a greater occurrence in higher mass progenitors, and may explain in part the preponderance of progenitor masses below 17\,\msun\ inferred from nebular spectra.
  }

   \maketitle

\section{Introduction}

Nebular-phase spectroscopy provides critical information on the properties of massive star explosions and type II supernovae (SNe). Originally cloaked by a massive, optically-thick ejecta, the inner metal-rich layers of the SN are revealed after about 100\,d as the H-rich material fully recombines and becomes transparent in the continuum. During this phase forbidden line emission, following collisional excitation and non-thermal excitation and ionization, is the dominant cooling process balancing \cofs\ decay heating.\footnote{In this study we use the electron energy balance \citep[e.g.,][]{Ost89_book, hm98} to discuss heating and cooling processes. In this equation (and in addition to the usual terms) the energy absorbed from radioactive decays appears as a heating term, while non-thermal excitation and ionization appear as coolant terms. Emission in H$\alpha$ does not appear as an explicit coolant. It is primarily produced via recombination but its strength is implicitly linked to non-thermal  processes through the coupling between the electron energy balance equation and the rate equations. In practice a significant fraction of the absorbed decay energy is emitted as H$\alpha$.} Such line emission conveys important information on the composition and the yields, the large-scale chemical mixing, the explosion geometry, and ultimately the progenitor identity (for a review, see \citealt{jerkstrand_rev_17}). In type II SNe, this is also the time when the original \nifs\ mass can be nearly directly (in the sense that it does not require any modeling of the SN radiation) extracted from the inferred SN luminosity (converting SN brightness to luminosity requires an estimate of the reddening and distance, and of the flux falling outside of the observed spectral range).

Nebular-phase spectroscopic modeling started in earnest with SN\,1987A, for which information has been gathered through continuous photometric and spectroscopic monitoring (for reviews, see for example \citealt{sn1987A_rev_90} and \citealt{mccray_rev_93}).  Numerous studies were focused on the nebular-phase radiation properties of SN\,1987A (see, for example, \citealt{fransson_neb_87A_87}; \citealt{KF92}; \citealt{li_87A_93}; \citealt{li_mccray_o1_92};  \citealt{li_mccray_ca2_93}; \citealt{li_mccray_he1_95}), and led to a refined understanding of the physical processes at play at these late times (for an extended discussion, see for example \citealt{fransson_chevalier_89}). The main conclusions from these studies is that, under the influence of \cofs\ decay heating, the various metal-rich and H-rich shells of the ejecta cool primarily through forbidden line emission of neutral and once-ionized species. Chemical mixing is inferred, although limited to large-scale macroscopic mixing and absent at the microscopic level. More advanced calculations have been performed since, with a more accurate and richer description of the ejecta composition, the atomic processes involved, and the atomic data employed \citep{KF98a,jerkstrand_87a_11}.  A more extensive analysis of SNe II-P and the physics relevant to nebular-phase modeling has been presented in a series of papers by \citet{jerkstrand_12aw_14,jerkstrand_04et_12,jerkstrand_ni_15}. This work was also made possible by the acquisition of nebular-phase spectra for nearby SNe, although even today the published sample remains limited to a handful of objects (see, for example, the dataset presented by \citealt{silverman_neb_17}).

An important conclusion from these works is that the \oidoub\ doublet flux, or its ratio with that of \caiidoub, can be used to constrain the progenitor mass. Applied to the existing dataset, nebular-phase modeling indicates initial progenitor masses below about  17\,\msun, with a noticeable absence of $20-25$\,\msun\ progenitors (e.g., \citealt{jerkstrand_ni_15}; see also \citealt{smartt_09}). A notable exception is the type II SN\,2015bs, for which a high mass progenitor of $20-25$\,\msun\ seems a plausible explanation for its unprecedented large \oidoub\ doublet flux \citep{anderson_15bs_18}.

With \cmfgen\ \citep{HD12}, we have conducted numerous simulations for type II SNe \citep{DH11_2p,d13_sn2p,lisakov_08bk_17,d18_fcl,d19_2pec} but generally limited these to the photospheric phase. At nebular times, our models match quite closely the observations of standard type II SNe like 1999em (see, for example, \citealt{d13_sn2p} and \citealt{silverman_neb_17}). With the treatment of non-thermal processes, \cmfgen\ predicts a strong H$\alpha$ line at nebular times (say 300\,d after explosion), while that line is absent in previous \cmfgen\ simulations in which non-thermal processes were ignored \citep{DH11_2p}. We shied away from the nebular phase because of our inability, in 1D, to treat chemical segregation satisfactorily. Indeed, in \cmfgen, we simultaneously enforce macroscopic and microscopic mixing, which conflicts with the properties of mixing seen in multidimensional simulations of core-collapse SN explosions (see, for example, \citealt{wongwathanarat_15_3d}). This reluctance is, however, questionable. First of all, our models match quite closely most of the observed SNe II at nebular times. Secondly, it is possible to introduce two different levels of mixing in our simulations. We may for example apply strong mixing of \nifs\ and daughter isotopes, but impose a very low level of mixing for all other species (in the current context of \cmfgen, this mixing is microscopic and macroscopic). This is not fully satisfactory in that we then microscopically mix \nifs, \cofs, and \fefs\ throughout the ejecta, but of these three species, Fe dominates at nebular times of 300\,d.\footnote{The importance of \nifs\ mixing is mitigated by the ability of $\gamma$-rays to travel some distance before being absorbed, so that the distribution of decay-power absorbed tends to be more extended than that of \nifs.} This Fe is already present at the microscopic level with a mass fraction of at least $\sim$\,0.001 throughout most of the ejecta -- this floor value is the solar-metallicity value (see Section~\ref{sect_caveats} for a discussion on the limitations of this statement). Hence, this approach is not so far from what takes place in Nature. We believe there is a clear interest in presenting a grid of simulations for SNe II at nebular times and exploring the sensitivity of SN radiation to our various ejecta properties. Such a grid has never been published.

In the next section, we present the numerical setup used for the ejecta and for the radiative transfer modeling. In section~\ref{sect_vturb}, we discuss the influence of the adopted Doppler width on the resulting ejecta and radiation properties. For most models in this paper, we adopt a low value of 2\,\kms. In Section~\ref{sect_ref_model}, we describe in detail the results for a SN II from a 20\,\msun\ star on the zero-age main sequence (ZAMS). This case, taken as a reference, is used to discuss the physics controlling nebular-phase spectra, the line formation process at nebular times, and to identify the predicted nebular lines in the optical and near-infrared ranges. Section~\ref{sect_o_over_ca} discusses the critical impact of the O/Ca ratio in the O-rich shell on the \oidoub\ doublet strength. Section~\ref{sect_henv} discusses the impact of the H-rich envelope mass on the nebular-phase properties, which is relevant for comparing the emission properties of SNe II arising from progenitors of lower and higher mass (say between a 12 and a $\geq$\,25\,\msun\ star). We then discuss the influence of the adopted \nifs\ mixing and of the \nifs\ mass on our nebular-phase spectra in sections~\ref{sect_ni_mix} and \ref{sect_var_mni}. While all previous simulations were performed at a SN age of 300\,d, section~\ref{sect_neb_evol} discusses the evolution of SN II properties from 150 to 500\,d after explosion. We discuss the influence of the progenitor mass (for a fixed \nifs\ mass and ejecta kinetic energy) on the nebular-phase spectra in section~\ref{sect_prog_mass}.  Section~\ref{sect_obs} presents a succinct comparison of our crafted nebular models with the observations of a few well observed SNe II at about 300\,d after explosion. Finally, we present our conclusions in section~\ref{sect_conc}.

\section{Context and numerical setup}

\subsection{Properties of massive star progenitors at core collapse: results from a grid of \mesa\ models}

     We have performed stellar evolution simulations with \mesa\ \citep{mesa1,mesa2,mesa3,mesa4} version 10108 for a 12, 15, 20, 25, 27, and 29\,\msun\ star on the ZAMS. All models were evolved until iron core formation and collapse (i.e., when the maximum infall velocity is 1000\,\kms). We assumed a solar metallicity mixture with $Z=0.014$ and no rotation. We used the {\tt approx21.net} nuclear network. We used the default massive star parameters in \mesa\ with the following exceptions. As in our previous works, we used a mixing length parameter of three (see \citealt{d13_sn2p}), which leads to smaller RSG surface radii. This modification was also used by \cite{mesa4} for their simulations of SNe II-P progenitors. We modified the parameters {\code{overshoot\_f0\_above\_nonburn\_core}} to be 0.001 instead of 0.0005 (same for {\code{overshoot\_f0\_above\_burn\_h\_core}}, {\code{overshoot\_f0\_above\_burn\_he\_core}} and {\code{overshoot\_f0\_above\_burn\_z\_core}}). We also modified {\code{overshoot\_f\_above\_burn\_z\_core}} and related quantities (counterparts for ``\code{h}" and ``\code{he}") to be 0.004. With these slight enhancements in overshoot, the stability of the code is improved during the advanced burning stages and the nucleosynthesis is somewhat boosted (in both respects, enhanced overshoot has a similar impact as introducing some rotation in the progenitor star on the ZAMS).  We used the ``Dutch" recipe for mass loss, with a scaling factor ``du" that we varied between zero (no mass loss) and one (we use factors of 0, 0.3, 0.6, 0.8, and 1.0). This allows us to gauge the influence of mass loss on the properties of the star (and thus to address the large uncertainty in mass loss rates) and, in particular, the stellar core at death.

     In about 50\% of the models produced with these parameters, the Si-rich and the O-rich shells merged during Si burning, producing a single Si-rich and O-rich shell with a nearly uniform composition (all elements are microscopically mixed). This has a considerable impact on the appearance of a core-collapse SN at nebular times \citep{fransson_chevalier_89}. Indeed, Ca then becomes abundant throughout the merged shell, and the \caiidoub\ doublet becomes the primary coolant, inhibiting the cooling through other lines and in particular the \oidoub\ doublet (see section~\ref{sect_o_over_ca}). Ca is also produced during the explosion, together with Si and similar elements. However, during the explosion there is no microscopic mixing of the Si-rich material with the O-rich shell because the mixing is exclusively macroscopic.

       In this study, we want to exclude such configurations to keep this complexity aside. The general wisdom so far has been to assume that these deep convective burning shells do not merge (see for example \citealt{fransson_chevalier_89}, but see discussion in sections~\ref{sect_o_over_ca} and \ref{sect_conc}). Hence, to prevent the merging of the Si-rich shell and the O-rich shell during Si burning with \mesa\, we set {\code{min\_overshoot\_q}} to 1 and {\code{mix\_factor}} to 0 when the central \iso{28}Si mass fraction first reaches 0.4. All simulations discussed in this section were produced in this manner and exhibit clearly distinct Si-rich and O-rich shells. The resulting model properties are given in Table~\ref{tab_shell_masses}, including initial and final masses, the main shell masses, and some ratios of mean mass fractions of important species within these shells. Multidimensional simulations of the last burning stages of massive stars prior to collapse are needed to determine the level of mixing, if any, of the Si-rich and O-rich shells (e.g., \citealt{meakin_arnett_07}; \citealt{couch_3d_presn_15}; \citealt{chatzopoulos_presn_16}; \citealt{mueller_3d_17}; \citealt{yoshida_presn_19}). A recent 3D simulation finds violent merging of the O and Ne shells in a star with an initial mass of 18.88\,\msun\  \citep{2020ApJ...890...94Y}. Convection is much more vigorous in 3D than 1D, and this leads to greater mixing.

\begin{table*}
\caption{Shell masses and abundance ratios in our grid of massive star models. The quantity \{X/Y\} corresponds to the total mass of element X over the total mass of element Y in the shell for the corresponding column. Numbers in parenthesis correspond to powers of ten.
\label{tab_shell_masses}}
\begin{center}
\begin{tabular}{c|c|c|c|c|cc|ccc|cc|c}
\hline
% N/He &    Mg/O & Ca/Si
 $M_{\rm init}$ & $Z$ & du & $M_{\rm final}$ & H-rich shell & \multicolumn{2}{c|}{He-rich shell}  & \multicolumn{3}{c|}{O-rich shell}  &
 \multicolumn{2}{c|}{Si-rich shell}  & Fe core  \\
\hline
 \,[\msun]      &    &     & [\msun]       & [\msun] & [\msun] & \{N/He\} &  [\msun]  & \{Mg/O\} & \{O/Ca\} & [\msun] & \{Ca/Si\} & [\msun] \\
\hline
 12 & 0.014 & 0.0 &12.0  &8.60  &1.33 & 7.64(-3)&0.37 &7.81(-2) & 1.53(-4)   &0.10 & 6.26(-2) &1.52   \\
 12 & 0.014 & 0.6 &10.3  &6.98  &1.39 & 5.85(-3)&0.36 &1.00(-1) & 1.63(-4)   &0.10 & 6.16(-2) &1.50   \\
 12 & 0.014 & 0.8 &9.7   &6.30  &1.41 & 7.38(-3)&0.38 &1.04(-1) & 1.95(-4)   &0.09 & 6.69(-2) &1.50   \\
 12 & 0.014 & 1.0 &9.2   &5.83  &1.44 & 2.37(-3)&0.29 &1.15(-1) & 2.62(-4)   &0.09 & 6.63(-2) &1.50   \\
 15 & 0.014 & 0.0 &15.0  &10.24 &1.60 & 6.43(-3)&1.14 &7.84(-2) & 3.13(-4)   &0.21 & 9.24(-2) &1.63   \\
 15 & 0.014 & 0.6 &12.4  &7.70  &1.59 & 5.44(-3)&1.15 &8.37(-2) & 2.56(-4)   &0.16 & 8.59(-2) &1.59   \\
 15 & 0.014 & 0.8 &11.3  &6.69  &1.58 & 5.76(-3)&1.02 &5.58(-2) & 3.94(-4)   &0.23 & 9.48(-2) &1.65   \\
 15 & 0.014 & 1.0 &10.3  &5.68  &1.66 & 2.44(-3)&1.07 &7.49(-2) & 1.95(-4)   &0.20 & 8.16(-2) &1.62   \\
 20 & 0.014 & 0.0 &20.0  &12.73 &2.19 & 2.31(-3)&2.91 &6.50(-2) & 1.92(-4)   &0.36 & 8.00(-2) &1.74   \\
 20 & 0.014 & 0.6 &15.4  &8.22  &2.17 & 2.40(-3)&2.89 &4.14(-2) & 1.38(-4)   &0.28 & 7.94(-2) &1.69   \\
 20 & 0.014 & 0.8 &13.4  &6.27  &2.11 & 2.67(-3)&2.77 &5.70(-2) & 1.65(-4)   &0.37 & 7.73(-2) &1.75   \\
 20 & 0.014 & 1.0 &11.6  &4.43  &2.13 & 2.75(-3)&2.86 &4.28(-2) & 1.28(-4)   &0.29 & 7.71(-2) &1.69   \\
 25 & 0.014 & 0.0 &25.0  &14.82 &2.72 & 2.29(-3)&4.89 &5.26(-2) & 3.05(-4)   &0.69 & 9.38(-2) &1.94   \\
 25 & 0.014 & 0.6 &17.7  &7.86  &2.70 & 2.39(-3)&5.02 &5.15(-2) & 1.60(-4)   &0.38 & 8.83(-2) &1.75   \\
 25 & 0.014 & 0.8 &15.0  &5.20  &2.59 & 2.50(-3)&4.69 &6.74(-2) & 2.31(-4)   &0.60 & 8.64(-2) &1.88   \\
 25 & 0.014 & 1.0 &12.6  &2.86  &2.53 & 2.46(-3)&4.78 &5.97(-2) & 2.01(-4)   &0.49 & 8.41(-2) &1.82   \\
 27 & 0.014 & 0.0 &27.0  &15.85 &2.72 & 5.97(-3)&5.82 &5.49(-2) & 3.00(-4)   &0.70 & 9.76(-2) &1.93   \\
 27 & 0.014 & 0.6 &18.6  &7.59  &2.80 & 2.26(-3)&5.68 &5.12(-2) & 2.35(-4)   &0.59 & 9.30(-2) &1.87   \\
 27 & 0.014 & 0.8 &15.9  &5.11  &2.69 & 2.34(-3)&5.50 &5.96(-2) & 2.68(-4)   &0.65 & 9.16(-2) &1.91   \\
 27 & 0.014 & 1.0 &13.7  &3.05  &2.67 & 2.27(-3)&5.48 &5.73(-2) & 2.47(-4)   &0.62 & 8.96(-2) &1.88   \\
 29 & 0.014 & 0.0 &29.0  &16.56 &2.98 & 2.49(-3)&6.82 &4.28(-2) & 3.05(-4)   &0.75 & 1.02(-1) &1.97   \\
 29 & 0.014 & 0.6 &20.3  &8.21  &2.83 & 2.28(-3)&6.73 &4.88(-2) & 1.91(-4)   &0.50 & 9.56(-2) &1.82   \\
 29 & 0.014 & 0.8 &17.7  &5.69  &2.89 & 2.13(-3)&6.67 &4.65(-2) & 1.89(-4)   &0.50 & 9.74(-2) &1.82  \\
 29 & 0.014 & 1.0 &15.7  &3.89  &2.83 & 2.22(-3)&6.47 &5.12(-2) & 1.96(-4)   &0.59 & 9.31(-2) &1.87  \\
\hline
\end{tabular}
\end{center}
\end{table*}

       With the adopted variation in mass-loss rate, the mass of the H-rich shell (or progenitor envelope) covers a large range from $\sim$\,3 up to $\sim$\,17\,\msun. Stellar winds in this mass range, metallicity, and adopted mass-loss rate recipes, do not peel the star all the way down to the He core so the He-rich shell and deeper metal-rich shells are not directly affected by mass loss. The mass of the He shell falls in a narrow range between about 1.3 to 3.0\,\msun. The mass of the O-rich shell follows a much greater variation, growing from around 0.3 in the 12\,\msun\ progenitor and rising to about 6.8\,\msun\ in the 29\,\msun\ progenitor, more than 20 times greater. The Si-rich shell follows a similar trend but increases by only a factor of six between 12 and 29\,\msun\ progenitors (i.e., it goes from about 0.1 to 0.6\,\msun). Excluding the artificial models that ignore mass loss, these shell masses do not vary much with the different scalings adopted (see third column in Table~\ref{tab_shell_masses}).

       In the O-rich shell, O is about 10$^4$ times more abundant than Ca, whose mass fraction in that shell is equal to the original metallicity in our \mesa\ simulations (set to solar in this work; see, however, section~\ref{sect_caveats}). If the Si-rich and O-rich shells were fully mixed, the same models would yield an O/Ca mass ratio in the range $60-200$, so typically 100 times smaller than the O/Ca in the unadulterated O-rich shell.

       The Mg to O mass ratio in the O-rich shell is comparable to the Ca to Si mass ratio in the Si-rich shell and is equal to about $0.05-0.1$. While the Mg to O mass ratio tends to decrease with main sequence mass the adopted mass-loss rate introduces a small scatter at a given progenitor mass. Finally, the N to He mass ratio in the He-rich shell is  around 0.005, with a scatter of about 50\,\% from low to high mass progenitors.

       These are representative composition properties for our set of \mesa\ simulations evolved with the network {\tt approx21.net}, which is routinely used for massive star explosions. We do not attempt in this study to further adjust the composition to reflect additional nuclear processes not accounted for by the network {\tt approx21.net}. Using state-of-the-art explosion models with a fully consistent composition computed with a huge network is straightforward but is delayed to our next study.

\subsection{A simplified description of core-collapse supernova ejecta}
\label{sect_setup}

Using a physical model of the explosion has the advantage of consistency. The ejecta structure and composition are determined by the laws that govern stellar evolution, stellar structure, and radiation hydrodynamics of stellar explosions. However, important insights can be gained by using an alternative approach in which the ejecta properties are defined analytically by using insights from progenitor and explosion models as a guide. One can then modulate such models without any effort. For example, one can vary the abundance profiles, the abundance ratios, the composition, or the chemical mixing without having to produce a new progenitor with a stellar evolution code and a new ejecta with a radiation hydrodynamics code for each new model.

In this work, we use the shell masses and representative abundance ratios of the main shells (as obtained in the grid of models presented in the previous section) to craft our ejecta composition. We assume that the ZAMS mass determines the final core properties. We further assume that the effect of mass loss is limited to trimming the star, thus causing no impact on the final core properties. This is largely corroborated by our \mesa\ simulations of single stars -- all models in the $10-25$\,\msun\ range start losing mass when reaching the RSG phase, and continue to lose mass until core collapse. In practice, mass loss can make a massive star evolve as if it was of a lower mass. However, this latter assumption only implies a shift between the ejecta properties and its corresponding ZAMS mass. Similar shifts can be caused by rotation, in the sense that a rotating star tends to produce core properties similar to those of a more massive but non-rotating star. These slight shifts in shell masses or final masses do not have any significance for the results discussed in this paper.

We generate ejecta models corresponding to 12, 15, 20, and 25\,\msun\ stars. The assigned masses for the He, CO, and Si cores are based on the values obtained in the stellar evolution calculations with \mesa\ and described in the previous section (see also Table~\ref{tab_prog_prop}). We add a fixed H-rich shell of 9\,\msun\ to the He-core (some type II SN models are also done by forcing the total ejecta mass to be 10\,\msun). Although the Fe-core mass covers the range $1.5-1.97$, we set it to 1.5\,\msun\ for all SN models.  Table~\ref{tab_prog_prop} summarizes the basic progenitor properties adopted in our grid of models. We adopt a fixed ejecta kinetic energy of 10$^{51}$\,erg and a fixed \nifs\ mass of 0.08\,\msun\ for all models unless otherwise stated (section~\ref{sect_var_mni} describes the influence of varying the \nifs\ mass on the SN radiation properties).

A further simplification is to limit the composition to the main elements witnessed in core-collapse SN spectra at nebular times. The radiative transfer process is the transformation of the decay power absorbed by the ejecta into escaping low-energy radiation. So, the gas properties are controlled by the coolants that balance the instantaneous decay heating (this balance is exact at nebular times). For the present exploration, we limit the composition to those atoms and ions that produce visible features in type II SN spectra at nebular times. In full simulations, left to a forthcoming study, we will include all species that play a role at some depth in the ejecta, even if those species produce no spectral mark.

We thus limit the composition of our ejecta to H, He, N, O, Mg, Si, Ca, Fe, Co, and Ni. We then prescribe the mass ratio for the dominant species of each shell, following the results from our \mesa\ calculations. The H-rich shell is made of 68\,\% H and 28\,\% He (all meant by mass). The He-rich shell is made of 99\,\% He and 0.5\,\% N (some models were inadvertently made with 2\,\% N). The O-rich shell is made of 90\,\% O and 10\,\% Mg. The Si-rich shell is made of 90\% Si and 10\,\% Ca. We prescribe a \nifs\ profile (function of the adopted mixing; see below). At nebular epochs, \nifs\ has decayed into a mix of \fefs, \cofs, and \nifs\ that have the same distribution in velocity space and a cumulative mass equal to the initial \nifs\ mass (\iso{58}Ni and \iso{59}Co are also included as part of the solar metallicity mixture).

In all shells we include the metals at their solar metallicity value whenever their atomic mass is greater than that of the dominant element in that shell (for example, we set Ca at the solar metallicity value in the O-rich shell but the O mass fraction is zero in the Si-rich shell). This property is motivated by our \mesa\ results (see Section~\ref{sect_caveats} for limitations of this choice). Computing models with a different metallicity is straightforward. Once all these mass fractions are set in each shell, we renormalize to unity at each depth (this introduces a change at the 10\,\% level). The mass ratios and shell masses define the default setup for all models. One can easily adjust a given abundance ratio or a shell mass to test the influence on the resulting SN observables.

Having set the distribution of elements in mass space, we then set the density versus velocity with the constraint that the kinetic energy should be 10$^{51}$\,erg. We primarily focus on one epoch, 300\,d after the explosion. However, we also investigate the evolution with time in section~\ref{sect_neb_evol}. The radius of each ejecta shell is directly set by the velocity for homologous expansion.

\begin{table}
\caption{Pre-SN progenitor properties for our model set. $M_{\rm H,e}$ refers to the H-rich envelope in the type II SN models. The He, CO, and Si core masses correspond to lagrangian masses.}
\begin{center}
\begin{tabular}{lccccc}
\hline
Model  &  $M_{\rm tot}$  &   $M_{\rm H,e}$  & $M_{\rm He,c}$  & $M_{\rm CO,c}$ &  $M_{\rm Si,c}$ \\
\hline
\multicolumn{6}{c}{Type II SN progenitors} \\
\hline
m12       &    12.4$^a$  &   9.0                &  3.4   &   1.9    & 1.6 \\
m15       &    13.5  &   9.0                &  4.5   &    2.8   &  1.75 \\
m20       &    15.9  &   9.0                &  6.9    &   4.9   &   2.0 \\
m25       &    18.3  &   9.0                &  9.3    &   7.1   &   2.3    \\
\hline
\end{tabular}
\end{center}
{\bf Notes:} $^a$ The type II SN model m12 has a total mass greater than the initial mass by 0.4\,\msun\ because of the adopted H-rich envelope mass. Its core properties are, however, compatible with the results from our \mesa\ grid. Such a model could arise from a merger \citep{menon_heger_87a_17}. 
\label{tab_prog_prop}
\end{table}%

\begin{figure}
\includegraphics[width=\hsize]{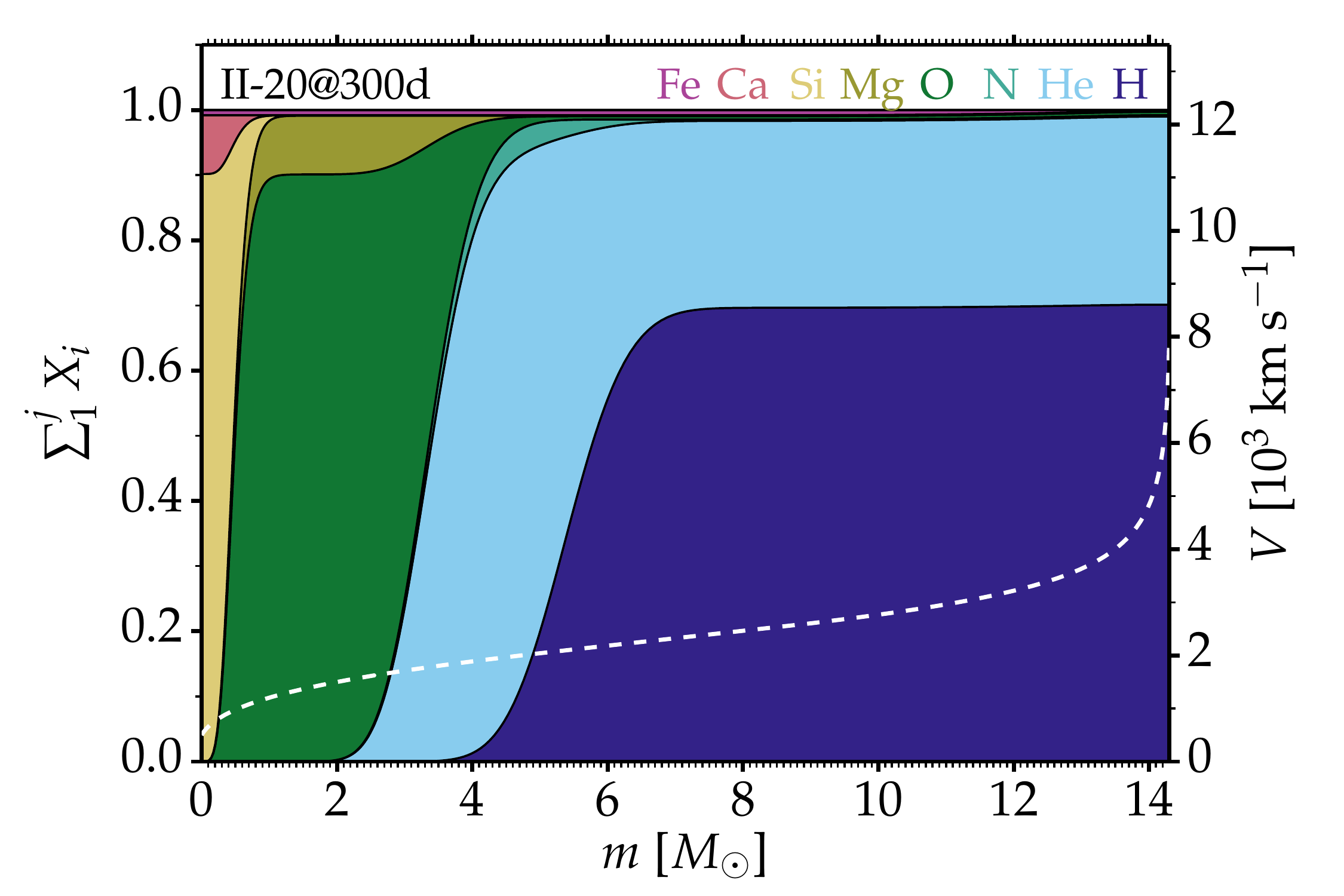}
\caption{Cumulative composition and velocity profiles versus Lagrangian mass in the ejecta of the SN II model based on the 20\,\msun\ ZAMS star. The sum for each plotted element $i$ includes the mass fractions for all plotted elements that have a lower atomic mass.
}
\label{fig_init_m20}
\end{figure}

The specification of the density versus velocity follows the approach of \citet{chugai_hv_07}.
The ejecta density distribution $\rho(V)$ is given by
\begin{equation}
    \rho(V)  = \frac{\rho_0}{1 + (V/V_0)^k}   \label{eq_rho}
\end{equation}
where $\rho_0$ and $V_0$ are constrained by the adopted ejecta kinetic energy
$E_{\rm kin}$, the ejecta mass $M_{\rm ej}$, and the density exponent $k$ (set to eight) through
\begin{equation}
M_{\rm ej} = 4 \pi \rho_0 (V_0 t)^3 C_m   \,\,   ;   \,\,
E_{\rm kin} = \frac{1}{2}\frac{C_e}{C_m} M_{\rm ej} V_0^2 \,\,  ,  \label{eq_rho_0}
\end{equation}
and where
\begin{equation}
C_m =  \frac{\pi}{k\sin(3\pi/k)}       \,\,   ;   \,\,
C_e   =  \frac{\pi}{k\sin(5\pi/k)} \,\,  .
\end{equation}
Finally, our guess for the initial temperature was a uniform value of 5500\,K. Under all conditions tested, this allows \cmfgen\ to converge steadily and robustly to the solution. With this choice, the converged temperature tends to be lower in the inner metal rich regions and higher as we progress in the H-rich shell, though the temperature typically stays within a few 1000\,K at most of this initial guess of 5500\,K. The final temperature profile (and the ionization etc) depends on many ejecta properties including the mass of \nifs\ and its distribution within the ejecta.

The composition that results from the above prescriptions presents a jump at the edge of each shell. We use a gaussian smoothing to broaden the composition profiles to mimic mixing. This is a tunable parameter so weak mixing (gaussian width of 100\,\kms) or strong mixing (gaussian width of 400\,\kms) were implemented. For greater freedom, the mixing of \nifs\ and other species is chosen to be independent.
The adopted initial profile for \nifs\ is of the form
\begin{equation}
     X(^{56}{\rm Ni}) \propto \exp(-Y^2)  \,\, {\rm with } \,\, Y = \frac{V -  V_{\rm Ni}}{\Delta V_{\rm Ni}}  \, ; \,  V \geq V_{\ \rm Ni} \,  \label{eq_vni}
 \end{equation}
and connects continuously to a constant \nifs\ mass fraction for $V < V_{\rm Ni}$. The normalization is set by the specified \nifs\ mass initially, which is 0.08\,\msun\ by default. By varying $V_{\rm Ni}$ and $\Delta V_{\rm Ni}$, we can enforce various levels of mixing of \nifs\ and its daughter elements, and therefore tune the spatial distribution of the absorbed decay power.  Figure~\ref{fig_init_m20} illustrates  the composition stratification for the type II SN ejecta corresponding to the 20\,\msun\ ZAMS mass.

\subsection{Radiative transfer modeling during the nebular phase with \cmfgen}

Using the ejecta configurations described above, we solve the non-local thermodynamic equilibrium (non-LTE) radiative transfer problem with \cmfgen\ \citep{hm98,DH05a,HD12}. Unlike in recent years, we here assume steady state so that a large number of simulations can be done independently without any knowledge of the previous evolution.  We use the fully relativistic transfer equations which are solved in a similar manner to the transfer equations discussed in \cite{HD12} and references therein. During the nebular phase, and at the typical velocities encountered in SNe, these give, for all practical purposes, the same results as found using the time-dependent transfer equation.

We treat non-thermal processes as per normal \citep{d12_snibc,li_etal_12_nonte}. We limit the radioactive decay to the \nifs\ chain. For simplicity, we compute the non-local energy deposition by solving the grey radiative transfer equation with a grey absorption-only opacity to $\gamma$-rays set to 0.06 $Y_{\rm e}$\,cm$^2$\,g$^{-1}$, where $Y_{\rm e}$ is the electron fraction. The model atoms included are: H\one\ (26,36), He\one\ (40,51), He\two\ (13,30), N\one\ (44,104), N\two\ (23,41), O\one\ (19,51), O\two\ (30,111), Mg\one\ (39,122), Mg\two\ (22,65), Si\one\ (100,187), Si\two\ (31,59), Ca\one\ (76,98), Ca\two\ (21,77), Fe\one\ (44,136), Fe\two\ (275, 827), Fe\three\ (83, 698), Co\two\ (44,162), Co\three\ (33,220), Ni\two\ (27,177), and Ni\three\ (20,107). The numbers in parentheses correspond to the number of super levels and full levels employed (for details on the treatment of super levels, see \citealt{hm98}).

Obviously, with the limited composition (and associated limited model atom), some lines will not be predicted in our simulations. This includes, for example, Na\one\,D so that the feature we predict around 5900\,\AA\ is primarily due to He\one\,5875\,\AA. Line blanketing associated with Ti\two\ is also neglected, although it tends to be much weaker at nebular times. Simulations with a detailed composition and the associated detailed model atom will be used in a forthcoming study.

For the initial conditions needed by \cmfgen, we assume a uniform temperature of 5500\,K throughout the ejecta, partial ionization of the gas, and all level populations are at their LTE value. Convergence to the non-LTE solution with a new temperature and electron density profile takes about 200 iterations, but only 12h of computing time for a Doppler width of 50\,\kms\ (see section~\ref{sect_vturb}). However, once a given model is converged, variants of that model (for example the same ejecta model but with a different \nifs\ mass or metallicity, etc.) can be computed quickly by adopting this converged model as an initial guess for the temperature, level populations, etc.

\subsection{Caveats}
\label{sect_caveats}

In our simplified approach, the properties of our adopted ejecta are not accurate nor consistent. We use the pre-SN composition of the main shells (which result from hydrostatic burning) to describe the SN ejecta composition (which results in part from explosive burning), with only one change associated with the presence of a \nifs-rich shell. In practice, this simplification amounts to introducing a \nifs-rich shell at the base of the ejected He-core material, leaving the composition of this pre-collapse He core unchanged.

This departs from what may occur in realistic explosions, whereby a large fraction of the Si-rich shell collapses into the compact remnant, while the \nifs-rich and Si-rich shells are recreated from explosive burning at the base of the O-rich shell. Strictly speaking, the yields from explosive and hydrostatic burning differ (see, e.g., \citealt{arnett_book_96}), but these differences are small to moderate. For the current study, the essence is simplification so that we can explore a variety of sensitivities of SN radiation to ejecta properties, focusing on the strength of the strongest lines, the influence of \nifs\ on Ca ionization and other ejecta properties. We have already performed simulations with a state-of-the-art ejecta composition (based on the ejecta models of \citealt{WH07}) and these do not yield critical differences that would impact the conclusions drawn in this paper (Dessart \& Hillier, in preparation).

Another limitation in our work is that the original shell compositions are based on \mesa\ simulations that are performed with a small nuclear network of only 21 isotopes. Although incomplete,  it is standard practice in the community for core-collapse SN simulations  since it captures the key physics while keeping the computing time small. It is well known that both resolution and network size alter the results of simulations of massive star evolution \citep{farmer_mesa_16} and explosion \citep{mesa3}, but not to an extent that would alter the results of the present study. For example, the s-process can lower the Ca and Fe abundance within the O-rich shell. The s-process is ignored by our small network and we find instead that the Ca and Fe mass fractions in the O-rich shell are essentially constant in our \mesa\ simulations. In the SN model s15 of \citet{WH07}, based on a 15\,\msun\ star initially, the s-process influences the Ca and Fe abundances so that they show a depression relative to solar by a factor of $5-10$ in the O/C shell, and by a factor of $2-3$ in the O/Ne/Mg shell. Ultimately, such features should be properly handled, but we ignore them for the time being. Since the Ca abundance is already 10000 times lower than the O abundance in the O-rich shell (in the absence of shell merging), a further reduction by a factor of a few  does not appear a critical change. When crafting our models, the same assumptions are applied to all ejecta models.  Our study focuses on trends, not on accurate quantitative assessments of line fluxes and ejecta yields. Slight offsets in abundances are therefore irrelevant for the conclusions we make.

\subsection{Set of simulations}

The parameter space that can be explored with our setup is large. One can change parameters in isolation and then try all possible permutations in a systematic way. We choose to be selective and vary parameters that reflect the range of possibilities consistent with our current understanding of stellar evolution and explosion physics. This includes the level of mixing of all species and the separate issue of \nifs\ mixing, as well as the influence of mass loss in peeling the star's envelope, yielding a pre-SN star with various amounts of H-rich material.

In the sections below we present results for simulations in which we varied the adopted Doppler width; the level of mixing either of all species or of \nifs\ or sometimes both; the mass of the H-rich envelope in the progenitor; the mass of \nifs; and the initial star mass. The mass fraction of elements within shells was also varied, in particular to test the influence of the O/Ca mass fraction ratio in the O-rich shell. We varied the metallicity from 0.1 to 2 times solar but found no influence on our results. For all simulations presented here, we set the metallicity to solar. Unless otherwise stated, all simulations adopt a SN age of 300\,d.

We do not provide a summary table of all models presented since the  models have many properties. Instead, in each section, we emphasize the given parameter that is varied while other parameters are kept identical. Our interest is in the influence of that varying parameter, not the numerous other parameters characterizing each individual model. In each section, the model nomenclature is clear enough to interpret the results and identify each model.

\begin{figure*}
\includegraphics[width=\hsize]{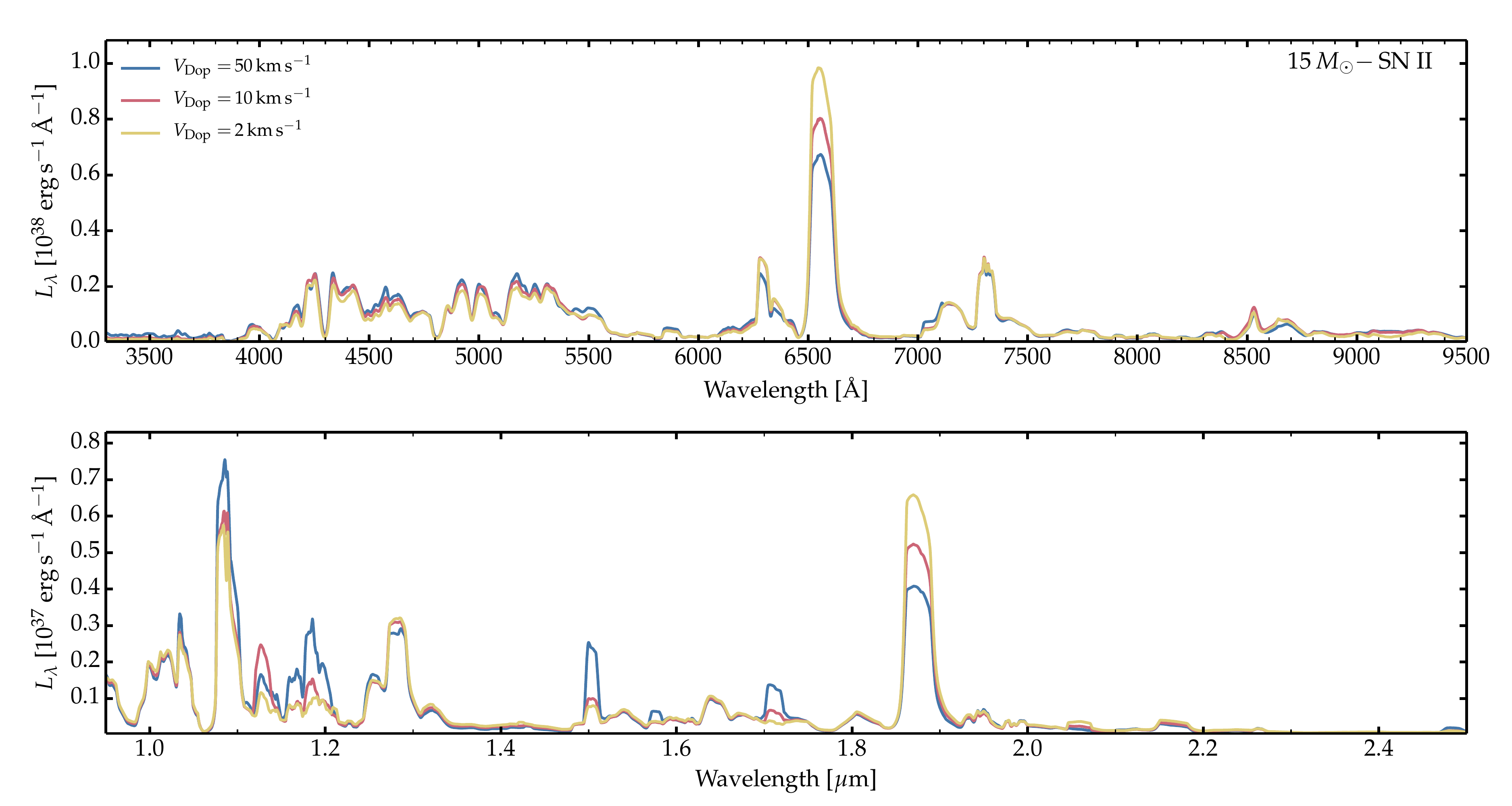}
\caption{Influence of the adopted Doppler width on the resulting optical and near-infrared spectra for a SN II model arising from a 15\,\msun\ progenitor. A weak mixing of all species is used except for \nifs\ for which we adopt a strong mixing. The increase in H$\alpha$ line flux with decreasing $V_{\rm Dop}$ is at the expense of a decrease in flux from a forest of Fe\two\ lines in the $4000-6000$\,\AA\ region. Other lines affected are He\one\,1.083\,$\mu$m, O\one\,1.129\,$\mu$m, Fe\one\ lines around 1.165\,$\mu$m, Mg\one\ lines at 1.183, 1.502, 1.577, and 1.711\,$\mu$m, and H\one\,1.875\,$\mu$m. Note that about 70\% of the total flux falls in the optical range.
\label{fig_vturb}
}
\end{figure*}

\section{Influence of the adopted Doppler width}
\label{sect_vturb}

For nebular phase radiative transfer modeling, most codes use the Sobolev approximation (e.g., \citealt{jerkstrand_87a_11}), which is equivalent to assuming an intrinsic line width of zero. In \cmfgen, this approximation is not used and all lines have a finite width.

Two terms control the intrinsic\footnote{Intrinsic in the sense that it applies in the comoving or gas frame, and is thus present even if the gas is as rest.} broadening of lines. The first one is associated with the thermal velocity of the corresponding atom or ion, and typically amounts to a few \kms\ at the low temperatures of SN ejecta at nebular times. This velocity width scales with $1/\sqrt{A}$, where $A$ is the atomic mass, so the line width of iron group elements is roughly a factor eight narrower than those of hydrogen at the same gas temperature. The other term is associated with turbulence. In our simulations we typically use the same broadening for all lines, and do this by ignoring the thermal contribution and by setting the turbulence to 50\,\kms. This choice is motivated by speed, and the need to have a ``reasonable'' number of frequency points (still of order 10$^5$) to cover the full spectrum from the far ultraviolet to the far infrared.

Tests in this study show that the adopted Doppler width influences the resulting spectra at  nebular times The choice of 0\,\kms\ implied by the Sobolev approximation is the opposite extreme and is probably not optimal since it prevents any line overlap (and hence interaction) within the Sobolev resonance zone. We adopted a value comparable to that implied by the thermal motions of the gas for IGEs.

In this work, we  first converged all models using a fixed Doppler width of 50\,\kms\ for all species. This produces a converged model that is pretty close to the true solution (the essential part of the solution being the converged temperature, ionization, and level populations). This converged model is then used as initial conditions for a new \cmfgen\ model in which a fixed Doppler width of 2\,\kms\ is used for all species (in cases where the impact on the gas properties was large, this reduction was done in a few steps).\footnote{We also run tests in which the Doppler width was determined based on the species atomic mass and a microturbulent velocity of 2\,\kms.  This is similar to using a Doppler width of $\sim$\,10\,\kms\ for H, 5\,\kms\ for He, and a few\,\kms\ for other species.}  The frequency grid is set so that all lines are resolved, irrespective of their widths. So, reducing the Doppler width of lines from 50 to 2\,\kms\ leads to an increase in the number of frequency points from 49,352 to 841,208 in the present simulations. This model takes much longer per iteration but it requires fewer iterations to converge since it starts with a better guess. With the lower Doppler width, the temperature and ionization are slightly modified and cause a sizable change in the strength of the strongest lines and the Fe\two\ emission forest around 5000\,\AA. The evolution can be nonlinear, in the sense that reducing the line Doppler width can yield a non-monotonic increase or decrease in the flux of \oidoub, H$\alpha$, or \caiidoub.

Figure~\ref{fig_vturb} shows the influence on the optical and near-infrared spectrum of changing the Doppler width from 50 to 10 and 2\,\kms\ in a strongly mixed ejecta model corresponding to a 15\,\msun\ progenitor. Reducing the Doppler width leads to a strengthening of H$\alpha$ and Pa\,$\alpha$ (and a weakening of the near-infrared Mg\one\ lines) which may have resulted from the increase of the electron density and the H ionization in the recombined H-rich layers of the ejecta. The same test in a model with weaker mixing would yield a different impact because the modified mixing has a strong influence on the emergent spectrum. So, for example, in a weakly mixed model, the reduction of the Doppler width leads to a reduction of He\one\ 7068\,\AA\ and Mg\one\ lines in the near infrared, a strengthening of the Ca\two\ lines in the optical, but has little impact on the H\one\ lines. It is thus important to realize that the adopted Doppler width can influence nebular line ratios.

The influence of line overlap on non-LTE processes is well documented. For example, overlap of an O\three\  line with He\two\ Ly\,$\alpha$ in planetary nebula leads to anomalous line strengths for several optical O\three\ lines \citep[e.g.,][]{Bow34_fluro,Ost89_book}. In some O stars the chance overlap of  Fe\four\ lines with the He\one\ resonance transition at 584\AA\ influences the strength of He\one\ singlet transitions in the optical (particularly those involving the 1s\,2p\,$^1$P$^{\rm o}$ level; \citealt{NHP06_HeI}). Another example, is the overlap of an O\one\ line with Ly\,$\beta$ which can, for example, influence the strength of O\one\ $\lambda$\,8446 \citep[e.g.,][]{Ost89_book}. As discussed by \cite{jerkstrand_04et_12}, a significant concern for modeling the nebular phase of Type IIP SNe is the overlap of some Fe\two\ and O\one\ lines with Ly\,$\alpha$ and Ly\,$\beta$. We verified the cause of the spectral changes by running test calculations which used a Doppler width of 50\,\kms\  and  modified model atoms for which we artificially reduced the oscillator strengths (effectively setting them to zero) of  Fe\two\ and O\one\ lines overlapping with  Ly\,$\alpha$ and Ly\,$\beta$.

\begin{figure*}
\includegraphics[width=\hsize]{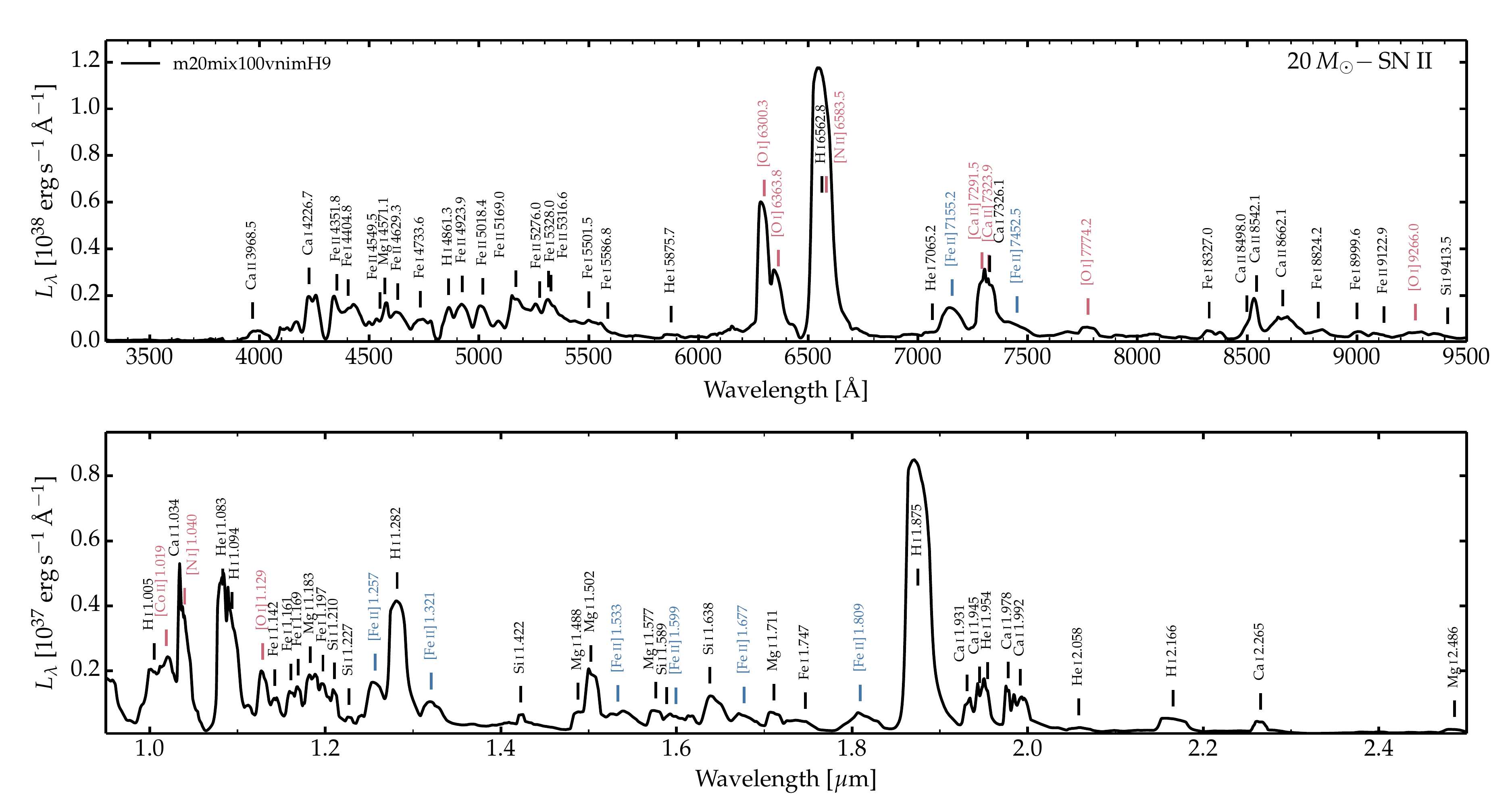}
\includegraphics[width=\hsize]{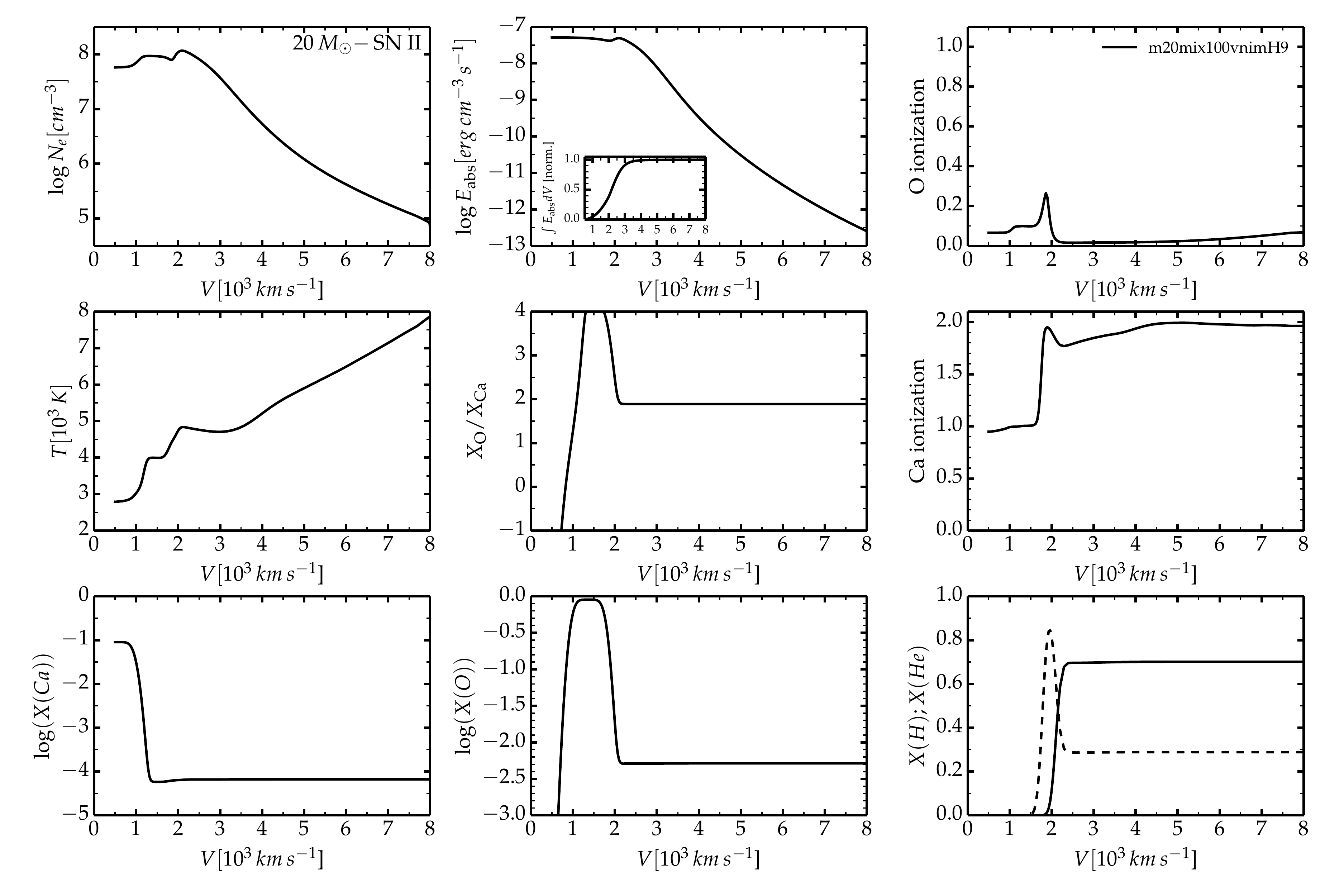}
\caption{Illustration of properties for the reference model  m20mix100vnimH9 discussed in section~\ref{sect_ref_model}. In this and similar figures, the top panels show the luminosity $L_\lambda$ (integrating over wavelength yields the bolometric luminosity) in the optical (upper) and the near infrared (lower). The line identifications are indicative only since in numerous cases multiple lines contribute (we give the primary component to most features). Forbidden lines are shown in red for all ions and atoms apart from Fe (shown in blue). The bottom panels show some ejecta properties computed with \cmfgen.
\label{fig_ref_model}
}
\end{figure*}

\section{Discussion of model results from a 20\,\msun\ type II SN model}
\label{sect_ref_model}

   We first describe the case of a type II SN ejecta from a 20\,\msun\ ZAMS star. The model is named m20mix100vnimH9, where ``mix100'' means that the gaussian smoothing uses a characteristic width of 100\,\kms\ (this mixing is applied to all species), ``vni'' means that the additional mixing applied to \nifs\ was strong ($V_{\rm Ni} =$\,2500\,\kms\ and $\Delta V_{\rm Ni}=$\,1000\,\kms; see Eq.~\ref{eq_vni}), and ``mH9'' means that the pre-SN progenitor had a 9\,\msun\ H-rich outer shell. The ejecta has a kinetic energy of 10$^{51}$\,erg and contains 0.08\,\msun\ of \nifs\ initially.

  A summary of results for this reference model is shown in Fig.~\ref{fig_ref_model}. The top two panels show the optical and near-infrared spectral properties. Throughout the paper, we show luminosities $L_\lambda$ (in erg\,s$^{-1}$\,\AA$^{-1}$) rather than scaled or normalized fluxes because its integral over wavelength yields the bolometric luminosity and thus the decay power absorbed by the ejecta. Variations in \nifs\ mass or $\gamma$-ray trapping efficiency will yield different brightnesses, which can be assessed from $L_\lambda$. The bottom panels show various properties of the ejecta, including the ionization level (electron density, O and Ca ionization), the temperature, the composition (for H, He, O, or Ca, and for the ratio of O/Ca, which is relevant for the \oidoub\ doublet strength in core-collapse SNe; see next section),  and the profile of the decay power absorbed.

\begin{figure*}
\includegraphics[width=\hsize]{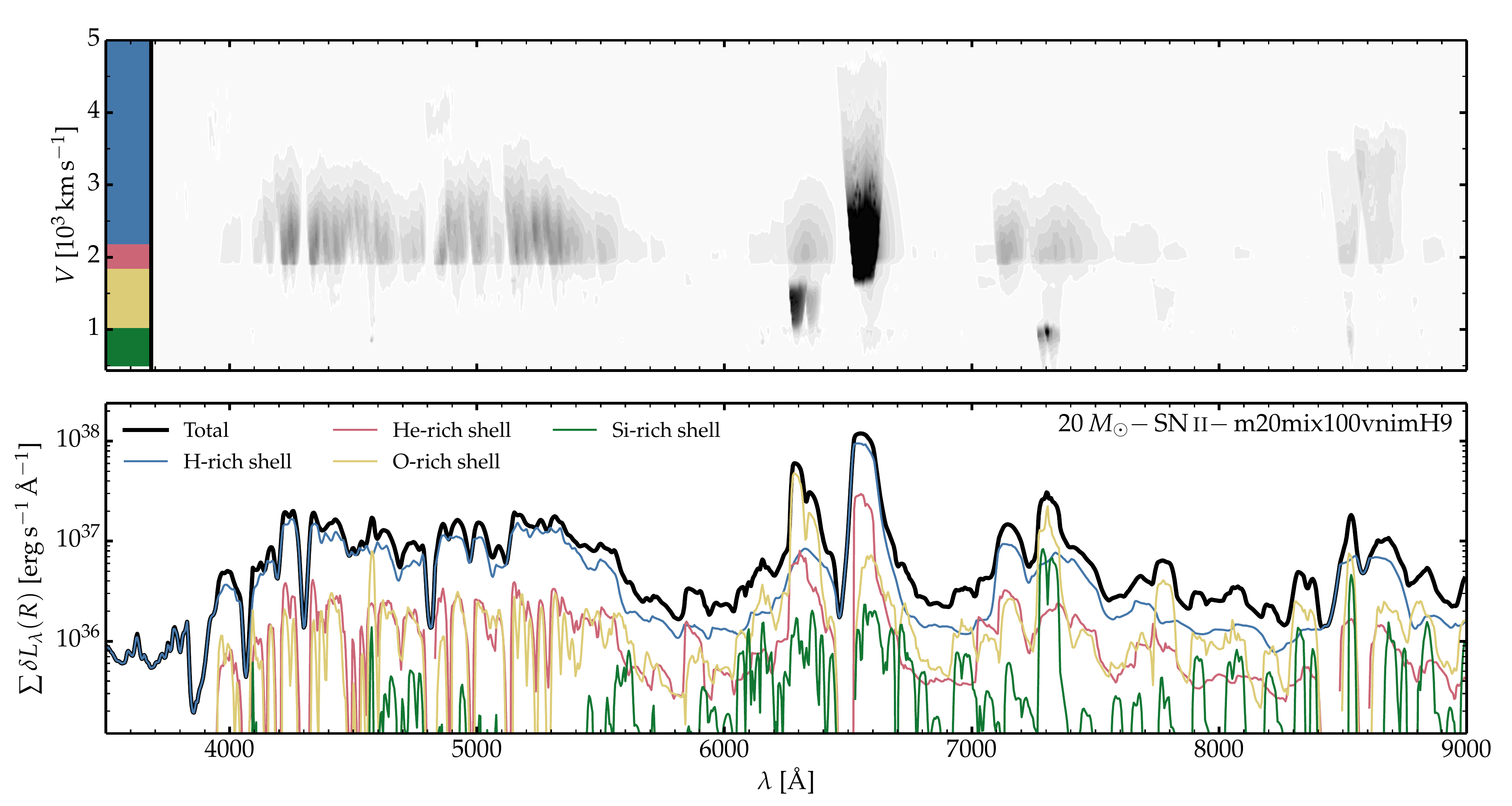}
\caption{Illustration of the spatial regions (here shown in velocity space) contributing to the emergent flux in a 20\,\msun\ type II SN model characterized by weak mixing for all non-IGE species, strong mixing for IGE species, and with a 9\,\msun\ H-rich envelope mass (model m20mix100vnimH9). The top panel shows the observer's frame luminosity contribution $\delta L_{\lambda,R}$ (Eq.~\ref{eq_dfr}; the map maximum is saturated at 20\,\% of the true maximum to bias against the strong H$\alpha$ line and better reveal the origin of the weaker emission) versus wavelength and ejecta velocity. The four contributing shells in this type II SN model are clearly seen (see vertical colored stripe at left), although the H-rich layers contribute most of the emergent flux. The bottom panel shows the emergent luminosity integrated over each shell, together with the total luminosity. In this model, the fraction of the total power emerging from the Si-rich, O-rich, He-rich and H-rich shells is 4\%, 20\%, 15\%, and 61\%, respectively. This illustrates the contribution to the emergent total flux from the main ejecta shells (this may also be inferred from the width of lines but complicated when multiple regions contribute or when there is line overlap). Note that with our approach, IGEs are present in all shells and thus Fe\one\ and Fe\two\ flux contribution is present throughout the ejecta. One can also see that the flux from the Si-rich shell is mostly radiated by Ca\one\ and Ca\two\ (together with Fe\one\ and Fe\two, rather than by the dominant species Si).
\label{fig_dfr}
}
\end{figure*}

The total radial electron scattering optical depth for this 300\,d-old ejecta is 0.4. Under such conditions, the ejecta no longer traps radiation so the energy balance is set by the distribution of decay power absorbed at that time in the ejecta and the reprocessing of this power by the gas in the form of low-energy, mostly optical, photons -- whatever is absorbed is instantaneously radiated and the balance is zero. Because of the strong \nifs\ mixing and some non-local energy distribution, the decay power is absorbed throughout the ejecta. Relative to the total energy absorbed, the Si-rich layers below 1000\,\kms\ get $\sim$\,20\,\%, the O-rich layers between 1000 and 1800\,\kms\ get $\sim$\,40\,\%, the He-rich layers between 1800 and 2100\,\kms\ get $\sim$\,10\,\%, and the outer H-rich layers receive the remaining  $\sim$\,30\,\%. Interestingly, the amount of radiation escaping to infinity and arising from these shells is very different, as shown in Fig.~\ref{fig_dfr}. In other words, there is a significant reprocessing of these low-energy photons by the ejecta.

\begin{figure}
\includegraphics[width=\hsize]{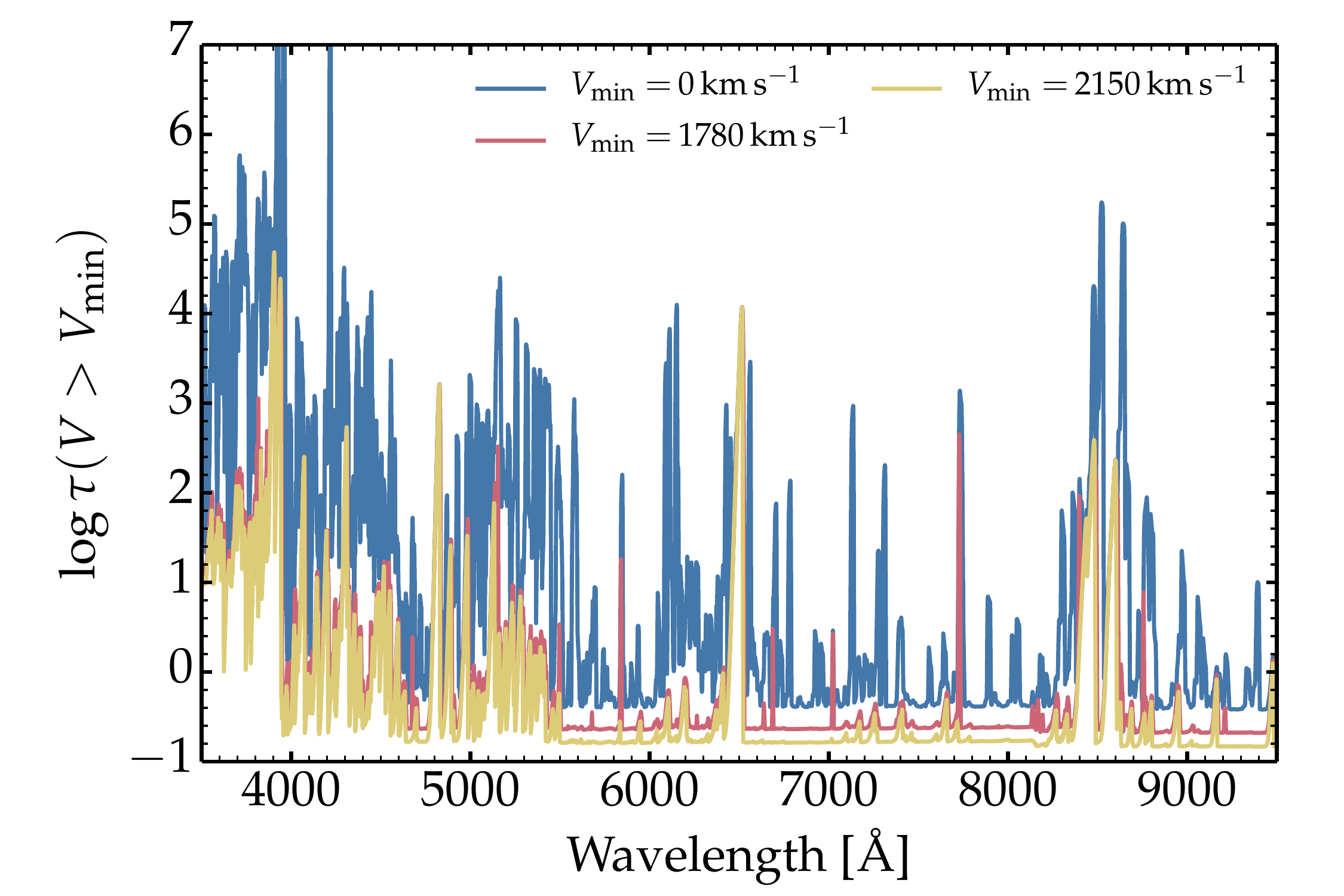}
\caption{Illustration of the total (i.e., accounting for all opacity sources) radial optical depth integrated from the outer ejecta boundary to the innermost boundary (i.e., the base of the ejecta; blue), to the He/O shell interface at 1780\,\kms\ (red), and to the He/H shell interface at 2150\,\kms\ (yellow) for the reference model m20mix100vnimH9.
\label{fig_tau}
}
\end{figure}

   Figure~\ref{fig_dfr} illustrates the spatial origin of the emerging flux and more specifically the quantity $\delta L_\lambda(R)$ -- it illustrates the location of the last interaction (not necessarily the location where the photon was originally emitted). The fractional contribution to the total flux at wavelength $\lambda$ from a narrow ejecta shell at radius $R$ is:
\begin{equation}
\label{eq_dfr}
\delta L_\lambda(R) = 8\pi^2\int \Delta z\
\eta(p,z,\lambda)\ e^{-\tau(p,z,\lambda)}\ p\ dp,
\end{equation}
\noindent
where $\Delta z$ is the projected shell thickness for a ray with impact parameter $p$, $\eta$ is the
emissivity along the ray at $p$ and $z$, and $\tau$ is the total ray optical depth at $\lambda$ (i.e., the integral is performed in the observer's frame) at the ejecta location $(p,z)$. Figure~\ref{fig_dfr} shows that the bulk of the flux emerges from the H-rich ejecta layers, which is 61\,\% of the total and thus about twice as much as deposited by $\gamma$-rays and positrons. The He-rich layers radiate 15\,\% of the total, compared to the 10\,\% of the decay power that they absorb. The O-rich (Si-rich) layers radiate 20\,\% (4\,\%) of the total, compared to the 40\,\% (20\,\%) of the decay power that they absorb. In the absence of optical depth effects, the power from each shell would be equal to the decay power absorbed in each corresponding shell. With a total electron-scattering optical depth of 0.4 at the SN age of 300\,d, the ejecta is not thin. It is not optically thick enough to cause radiation storage and diffusion with a sizable delay, but it is optically thick enough to cause strong reprocessing of UV and blue optical photons, which benefits the emission from the outer layers in this model. Such optical depth effects and their wavelength-dependent impact are illustrated in Fig.~\ref{fig_tau}.

   Figure~\ref{fig_dfr} also illustrates the complicated line formation process at nebular times. It helps in solving the problem of line overlap since a feature may appear broad with a narrow peak because it forms throughout the ejecta (from small to large velocities) or because it forms in a single narrow shell but overlaps with adjacent lines. Let us consider the formation of the strongest optical lines, which are mostly forbidden. The \oidoub\ doublet forms primarily in the O-rich shell (yellow line in Fig.~\ref{fig_dfr}), with a small contribution at large velocity from the H-rich shell (blue line) and a small contribution at low velocity from the He-rich shell (red line). This contribution from the He shell arises from our imposed mixing and is confined to the region at the interface between the two shells (Fig.~\ref{fig_init_m20}).

\begin{figure}
\includegraphics[width=\hsize]{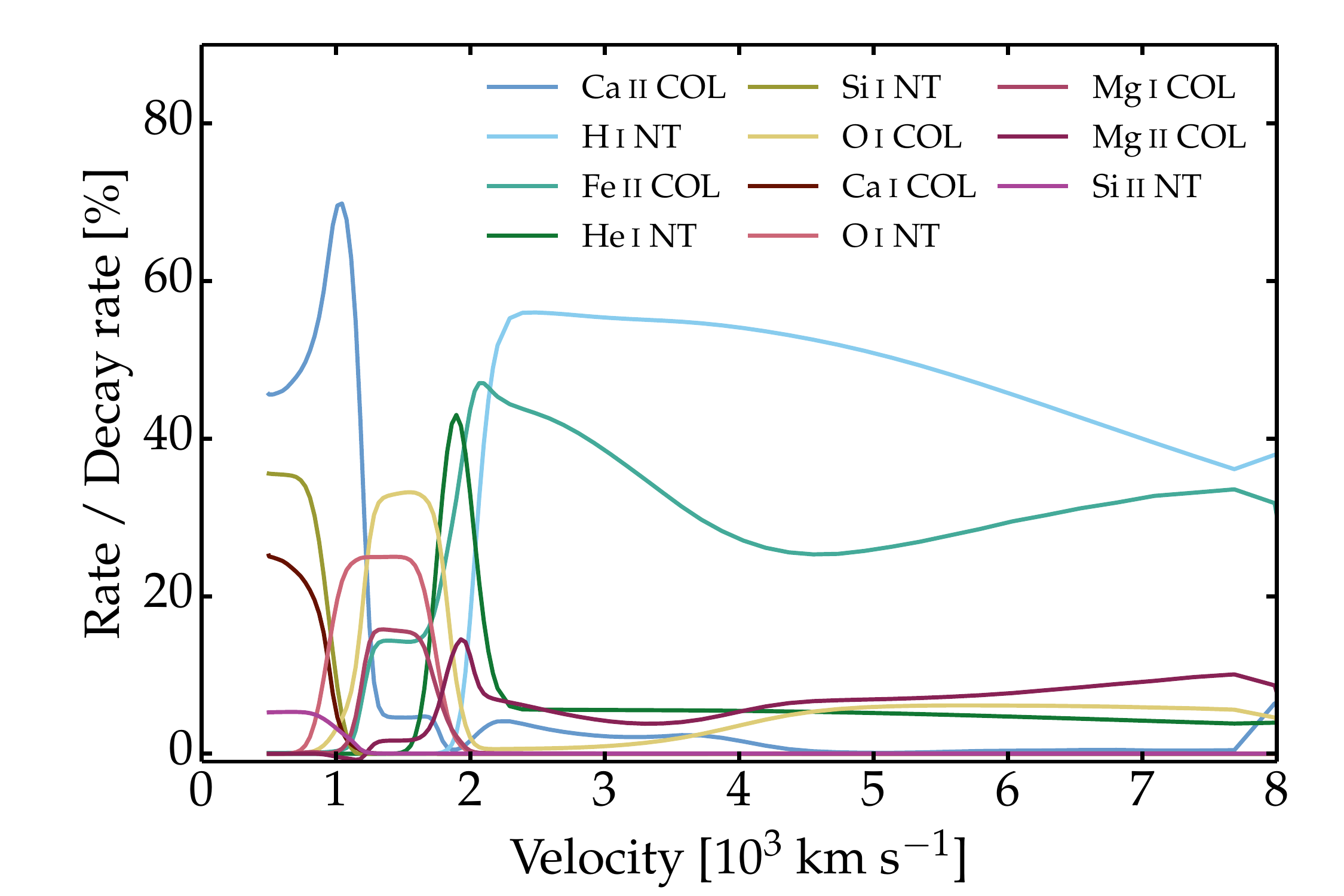}
\caption{Illustration of the main cooling processes balancing the radioactive decay heating at all ejecta depths in the reference model m20mix100vnimH9.  We show each dominant cooling rate (stepping down from the rate having  the large peak value at any depth) normalized to the local heating rate. The term ``NT'' stands for non-thermal processes (in this context non-thermal excitation and ionization) and ``COL" stands for collisional processes (i.e., collisional excitation).
\label{fig_cool}
}
\end{figure}

   The \caiidoub\ doublet forms throughout the ejecta -- from the outer part of the Si-rich shell and the inner part of the O-rich shell, and from the H-rich shell (very broad component). The \caiidoub\ doublet emission extends up to 7400\,\AA\ because of the contribution from the H-rich shell but also because of the influence of electron scattering in the fast moving H-rich layers. The emission further to the red is due to Fe\two\ (strongest component at 7452\,\AA). The He-rich shell, which is nearly exclusively He with a small amount of N, contributes mostly through the emission of N\two\,6548\,\AA, but this emission overlaps with the strong and broader H$\alpha$ line seen at all times in SNe II. The N\two\ identification is thus nontrivial in SNe II, but it is observed and explained in some SNe IIb \citep{jerkstrand_15_iib}. Below 6000\,\AA, most of the flux emerges from the H-rich layers and stems mostly from Fe\one\ and Fe\two\ line emission (with a mix of permitted and forbidden transitions). The strongest Fe\two\ line is an isolated forbidden line at 7155.2\,\AA. There is a weak Mg\one\,4571\,\AA\ line from the O-rich shell (as well as the O\one\, 7774\,\AA\ further to the red). The nucleosynthetic signatures of the explosion and the pre-SN evolution are thus quite limited, with the main features being  \oidoub\ and \caiidoub. A more extensive list of line identifications is provided in the upper panels of Fig.~\ref{fig_ref_model}.

  In the near infrared, the synthetic spectrum shows lines from the Paschen series of H\one, He\one\,1.083\,$\mu$m, O\one\,1.129\,$\mu$m, Mg\one\,1.488, 1.502 and 1.711\,$\mu$m, Si\one\ at 1.227 and 1.422\,$\mu$m, numerous Ca\one\ lines just short of 2\,$\mu$m,  as well as numerous Fe\one\ and Fe\two\ lines spread throughout the near infrared. These identifications are often ambiguous since most features are a composite of different lines.

\begin{table*}
\begin{center}
\caption{Atomic properties associated with the \oidoub\ and \caiidoub\ doublet transitions.
\label{tab_o1_ca2}
}
\begin{tabular}{|l|l|l|c|c|c||c|}
\hline
Species & Transition & Levels & $\lambda$   & $A_{\rm ul}$   & $\Upsilon$(T$_4$=0.5) & N$_e$(crit)(T$_4$=0.5)\\
              &                     &            &   [\AA]          &    [s$^{-1}$]       &           [s$^{-1}$]       &   [cm$^{-3}$]       \\
\hline
%O\,{\sc i} & $^3$P$_2$-$^1$D$_2$ & 1-4 & 6300.3 & $6.50 \times 10^{-3}$ & 0.146(T$_4$/0.5) & $2.4 \times 10^6$ \\
%O\,{\sc i} & $^3$P$_1$-$^1$D$_2$ & 2-4 & 6363.8 & $2.10 \times 10^{-3}$ & 0.146(T$_4$/0.5) & $2.4 \times 10^6$ \\
%Ca\,{\sc ii} & $^2$S$_{1/2}$-$^2$D$_{3/2}$ & 1-2 & 7323.9 & 0.795 & 4.6 &  $ 6.3  \times 10^6$\\
%Ca\,{\sc ii} & $^2$S$_{1/2}$-$^2$D$_{5/2}$ & 1-3 & 7291.5 & 0.903 & 6.8 & $ 6.3  \times 10^6$ \\
%
O\,{\sc i} & $^3$P$_2$-$^1$D$_2$ & $1-4$ & 6300.3 & $ 5.096\times 10^{-3}$ &  &     \\
O\,{\sc i} & $^3$P$_1$-$^1$D$_2$ & $2-4$ & 6363.8 & $ 1.639\times 10^{-3}$ &  0.124(T$_4$/0.5)\tablefootmark{a} & $2.2 \times 10^6$ \\
O\,{\sc i} & $^3$P$_0$-$^1$D$_2$ & $2-4$ & 6391.7 & $ 7.230 \times 10^{-7}$ &   &  \\
Ca\,{\sc ii} & $^2$S$_{1/2}$-$^2$D$_{3/2}$ & $1-2$ & 7323.9 & 0.795 & 4.58 &  $ 5.8  \times 10^6$\\
Ca\,{\sc ii} & $^2$S$_{1/2}$-$^2$D$_{5/2}$ & $1-3$ & 7291.5 & 0.802 & 6.79  & $ 5.8  \times 10^6$ \\
\hline
\end{tabular}
\end{center}
\tablefoot{
\tablefoottext{a}{The value listed is the total collision strength for the $^3$P to $^1$D transition.}
}
\end{table*}

One aspect that controls the radiative properties of the nebular phase spectrum is the ionization (shown for O and Ca in the bottom panels of Fig.~\ref{fig_ref_model}). Here, the mass density is constant below about 2000\,\kms\ so the variations of the electron density below 2000\,\kms\ reflect the change in ionization and mean atomic weight as we proceed through the He-rich shell, the O-rich shell, and the Si-rich shell. Oxygen is primarily neutral in the O-rich shell but Ca is once ionized in the Si-rich shell. The jump in ionization and temperature in the He-rich shell arises because the dominant species in that shell (i.e., He) is a poor coolant.

  At all depths in the ejecta, the heating source is radioactive decay. However, because of the stratification in composition and the range of densities between inner and outer ejecta, the sources of cooling vary drastically with depth. In the H-rich layers, we find that the cooling is done through non-thermal excitation and ionization (primarily in association with H\one, and by a factor of ten weaker with He\one), and collisional excitation of Fe\two\ (and to a lesser extent Mg\two\ and O\one). In the He-rich shell, the cooling is done through collisional excitation of Fe\two\ and non-thermal processes tied to He\one. In the O-rich shell, cooling arises primarily from collisional excitation of O\one\ and non-thermal processes tied to O\one,  and then from collisional excitation of Mg\one\ and Fe\two. In the Si-rich shell, the cooling is done primarily through collisional excitation of Ca\two\ as well as Ca\one, plus non-thermal processes tied to Si\one. Figure~\ref{fig_cool} illustrates these various cooling components at different ejecta depths in this reference model m20mix100vnimH9.

 The above results need to be considered in view of our assumptions. First, the ejecta composition is limited to H, He, N, O, Mg, Si, Ca, and the \nifs\ decay chain elements so the lines present in the models are by design limited to these elements (in their neutral or once ionized state). The second aspect is that we adopt a strong mixing of \nifs\ with a weak mixing of the other (lighter) elements. This gives some preference to the intermediate and high velocity layers of the ejecta (although we have seen that the outer layers do a lot of reprocessing of the radiation emitted from the inner layers). Finally, because of the weak mixing of species other than \nifs, the metal-rich layers retain an onion-like shell structure. With macroscopic mixing only, material from all shells in the pre-SN star would coexist at a given velocity. The reprocessing of radiation from the inner ejecta by the various lumps of material would be much more complex than in the present configuration of shells of distinct composition stacked on top of each other.

\section{Importance of the Ca mass fraction in the O-rich shell}
\label{sect_o_over_ca}

Converting line strengths into abundances of the associated ion and species is one of the principal goals of nebular phase spectroscopy for any type of SN. There are however many caveats associated with this task. First, the radiation emitted by any ion is very dependent on the atomic physics of that ion. Second, it depends on the efficiency  of other species that are radiating  from the same region. Third, it depends on where the radioactive decay power emitted is absorbed by the ejecta. Hence, a fundamental aspect of nebular line emission is how the \nifs\ is mixed through the ejecta and how far the emitted $\gamma$-rays can travel in the ejecta.

Having determined the non-local distribution of this decay power and which fraction of the total is shared between each dominant ejecta shell (H-rich shell, He-rich shell, O-rich shell, and the Si-rich shell), the radiation emitted in each shell will occur through the lines that have the strongest cooling power. The cooling efficiency of a given forbidden line depends of course on the abundance of the corresponding ion (hence a function of mass fraction and ionization level), but also on atomic properties of the ion levels (oscillator strength, critical density, etc.). In practice, type II SN ejecta exhibit a wide disparity in ionization, composition, and density, and the various constituents have widely different atomic properties so that even trace elements can dominate the cooling of the gas (the same holds in H\two\ regions, whose cooling is controlled to a large extent by emission in lines of N\two, O\two\ and O\three\ while the dominant species are instead H and He; \citealt{Ost89_book}).

Below, we discuss the case of the doublet forbidden transitions \oidoub\ and \caiidoub\ because they are the strongest optical lines in type II SN spectra at nebular times (if we exclude H$\alpha$) and also because they are routinely used to set constraints on nucleosynthesis and progenitor properties. We then present simulations in which various amounts of Ca are introduced into the O-rich shell. These simulations show the strong impact Ca can have on O\one\ line emission. We then discuss the implications and comment on previous work.

\subsection{The cooling power of \oidoub\ and \caiidoub}

Let us consider a small gas volume of uniform composition with a gas temperature $T$ (or $T_4$ when expressed in units of 10$^4$\,K) and electron density $N_{\rm e}$. Below, we study how the emissivity of (or the cooling rate associated with) the \oidoub\ doublet compares with that of the \caiidoub\ doublet. We assume optically-thin emission in this simplified analysis. This approximation may not hold strictly, in particular for large Ca densities.

The collisional de-excitation rate per unit volume per unit time from the upper level ``u" to the lower level ``l" is given by
\begin{equation}
N_{\rm u} C_{\rm ul}= {8.67 \times 10^{-8} \over \sqrt{T_4} } {\Upsilon_{\rm lu} \over g_u} N_{\rm e} N_{\rm u} \, ,
\end{equation}
where $N_{\rm u}$ is the upper level population and $\Upsilon_{\rm lu}$ is the effective collision strength for the transition (data for collision strengths are taken from \cite{mendoza_83} for O\,\one\ and from \cite{MBB07_CaII_col} for Ca\,\two; energy levels are from NIST, while  $A_{\rm ul}$ values for O\one\ are from \cite{Ost89_book}
and those for Ca\two\ are similar to those in \cite{1969SoPh...10..311L}. \footnote{The values listed in the table are those used in the present calculations. Experimental lifetime measurements, and  theoretical calculations, suggest that a better estimate for A for the Ca\two\ transitions is 1.1 \citep[see ][]{MBB07_CaII_col}. The value listed in the NIST database is 1.3 \citep{NIST571}. It comes from a calculation by \cite{1951ApJ...114..469O} and has an indicated error of greater than 50\%. The O\one\ values are slightly lower than those in the  NIST database \cite{NIST571} ($5.6 \times 10^{-3}$ and $1.8 \times 10^{-3}$; $E < 7$\%) however  \cite{1988JPhys..49..129B} suggest even higher values ($6.7 \times 10^{-3}$ and $2.3 \times 10^{-3}$). While the adoption of  different cross-sections will (slightly) alter  line strengths, they will not affect any of the conclusions made in this paper.}

 At the critical density, this rate is equal to $N_u A_{ul}$. Thus
\begin{equation}
N_{\rm e}{\rm(crit)}= 1.16 \times 10^7 {g_u A_{ul} \sqrt{T_4}\over \Upsilon_{lu} }  \,\, {\rm cm}^{-3} $$  \, .
\end{equation}
For O\,{\sc i} we sum the $A_{ul}$ values for the two transitions, to get a critical density. For Ca II, we treat the
upper level as a single level ($A=\sum g_u A_{ul}/\sum g_u$, and $\Upsilon =\sum \Upsilon_{lu}$). In the following, we assume O\,{\sc i} and Ca{\, \sc ii} are the dominant ionization stages, and that levels other
than those listed above can be neglected. We use the atomic properties listed in Table~\ref{tab_o1_ca2}.

We may consider  two limiting cases for the strength of these forbidden lines, corresponding to electron densities much higher or much lower than the critical density of the transition. \\

\noindent
Case 1: $N_{\rm e} \gg N_{\rm e}$(crit).

In this case we can assume the upper level is in LTE relative to the ground state. Thus, the O\one\ line emissivity is given by

\begin{equation}
\eta({\rm O\,{\textsc i}})  =   h \nu_{ul} N_{u} A_{ul}  = \frac{g_u}{g_l} \,h \nu_{ul}   A_{ul} N_{\rm O} \,\exp\left({- \frac{h\nu_{ul}}{kT}}\right )  ,
\end{equation}
or
\begin{equation}
 \eta({\rm O\,{\textsc i}})  = \frac{5}{9} \,h \nu_{ul}   A_{ul} N_{\rm O} \,\exp\left({-1.4388 \over \lambda(\mum)T_4 }\right ).
\end{equation}

\noindent Similarly, for the Ca\two\ line ($g_u/g_l=10/2$), we have
\begin{equation}
\eta({\rm Ca\,{\textsc{ii}}})  = 5 \,h \nu_{ul}   A_{ul} N_{\rm Ca} \,\exp\left({-1.4388 \over \lambda(\mum)T_4 }\right ).
\end{equation}

\noindent Integrating these line emissivities over the O-rich shell volume yields the line luminosities $L$.
Taking the ratio we obtain,

\begin{equation}
\frac{{\rm L}({\rm O\,{\textsc i}})}{{\rm L}({\rm Ca\,{\textsc{ii}}})}= 1.08 \times 10^{-3} {N_{\rm O}\over N_{\rm Ca}}  \exp\left({-0.309 \over T_4}\right).\label{eq1}
 \end{equation}

\noindent
Case 2: $N_{\rm e} \ll N_{\rm e}$(crit).

In this case, the line emissivity is

\begin{equation}
\eta={8.67 \times 10^{-8} \over \sqrt{T_4} } {\Upsilon_{lu} \over g_l} N_e N_l h \nu_{ul} \exp\left({-1.4388 \over \lambda(\mum)T_4 }\right )
\end{equation}

\noindent
since every excitation gives a line photon. Thus, the ratio of line luminosities is now

\begin{equation}
\frac{{\rm L}({\rm O\,{\textsc i}})}{{\rm L}({\rm Ca\,{\textsc{ii}}})}= 2.81 \times 10^{-3}  {N_{\rm O}\over N_{\rm Ca}}  \exp\left({-0.309 \over T_4}\right) \label{eq2} \, .
 \end{equation}
In this case we could add a factor like $(T_4/0.5)^{0.5}$ since the O\one\ collision strength grows faster than that of Ca\two.  A factor of 2.5 (i.e., 40/16) is introduced if we express Eqs.~\ref{eq1} and \ref{eq2} with mass fractions.

In the optically-thin limit, the luminosity contrast between these two lines would be huge if Ca was as abundant as O in the O-rich shell (in practice, line optical depth effects would reduce this Ca\,\two\ emission). Equations~\ref{eq1} and \ref{eq2} show that the Ca\two\ line would dominate over the O\one\ line if the Ca abundance is at least $\sim$\,1\% of the O abundance in the O-rich shell. This situation does not occur if Ca has the solar abundance in the O-rich shell, but the merging of the Si-rich and O-rich shells may change this. The impact would depend on the mass of the Si-rich and O-rich shells in the pre-SN star (and prior to merging).  The shell masses grow with the progenitor mass, with a $2-3$ times faster increase for the O-rich shell mass.

\begin{figure*}
\includegraphics[width=\hsize]{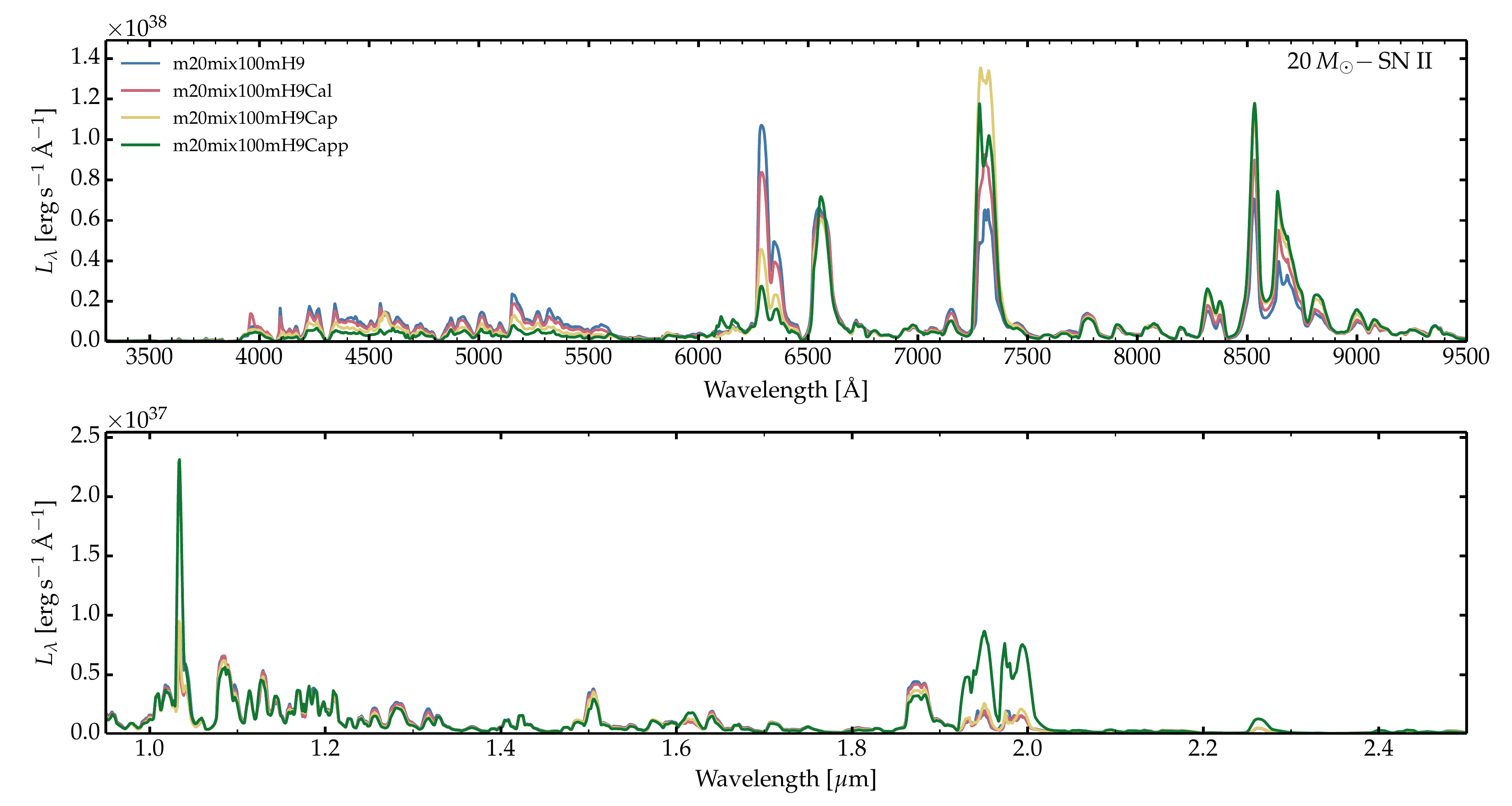}
\includegraphics[width=\hsize]{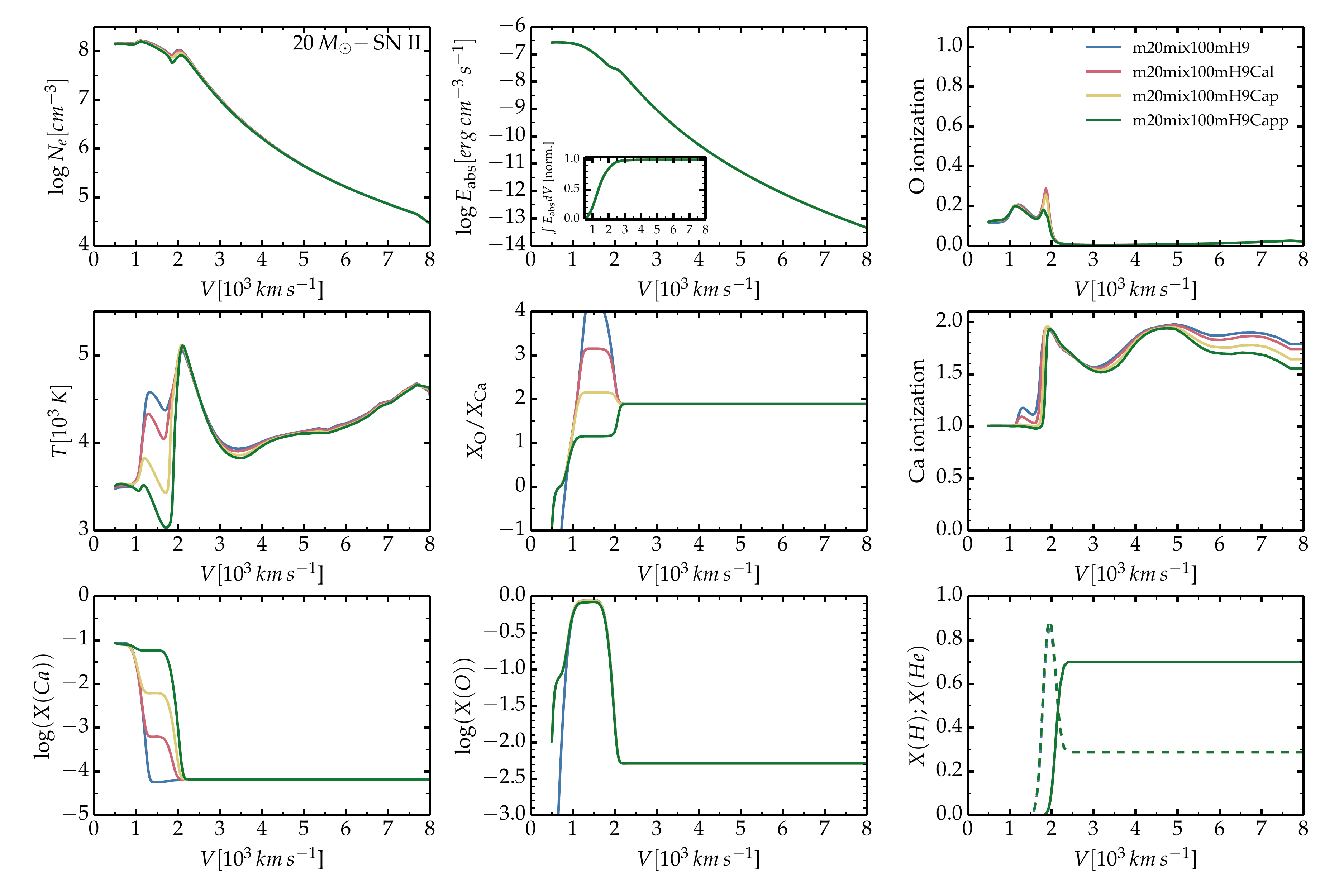}
\caption{Similar to Fig.~\ref{fig_ref_model}, but now showing variants of model m20mix100mH9 in which the Ca mass fraction is progressively increased in the O-rich shell by powers of ten and starting from about $7\times 10^{-5}$, the solar metallicity value (the Ca mass fraction profile is shown in the bottom left panel). Notice the dramatic reduction on the [O\one]/[Ca\two] line ratio as the Ca abundance is increased. Even in the richest Ca model, the Ca mass is still less than 10\% of the O mass.
\label{fig_Ca_in_Oshell}
}
\end{figure*}

\subsection{The influence of O/Ca in the O-rich shell: numerical simulations with \cmfgen}

Using as a reference the model m20mix100mH9 produced with the prescriptions stipulated in section~\ref{sect_setup}, we produce three other models in which the Ca mass fraction in the O-rich shell is raised from its solar value of $6.4\times 10^{-5}$ to 0.0007 (model suffix Cal), 0.007 (suffix Cap), and finally 0.07 (suffix Capp). The \cmfgen\ results for both the optical and near-infrared radiation and the gas properties are shown in Fig.~\ref{fig_Ca_in_Oshell}.

\begin{figure}
\includegraphics[width=\hsize]{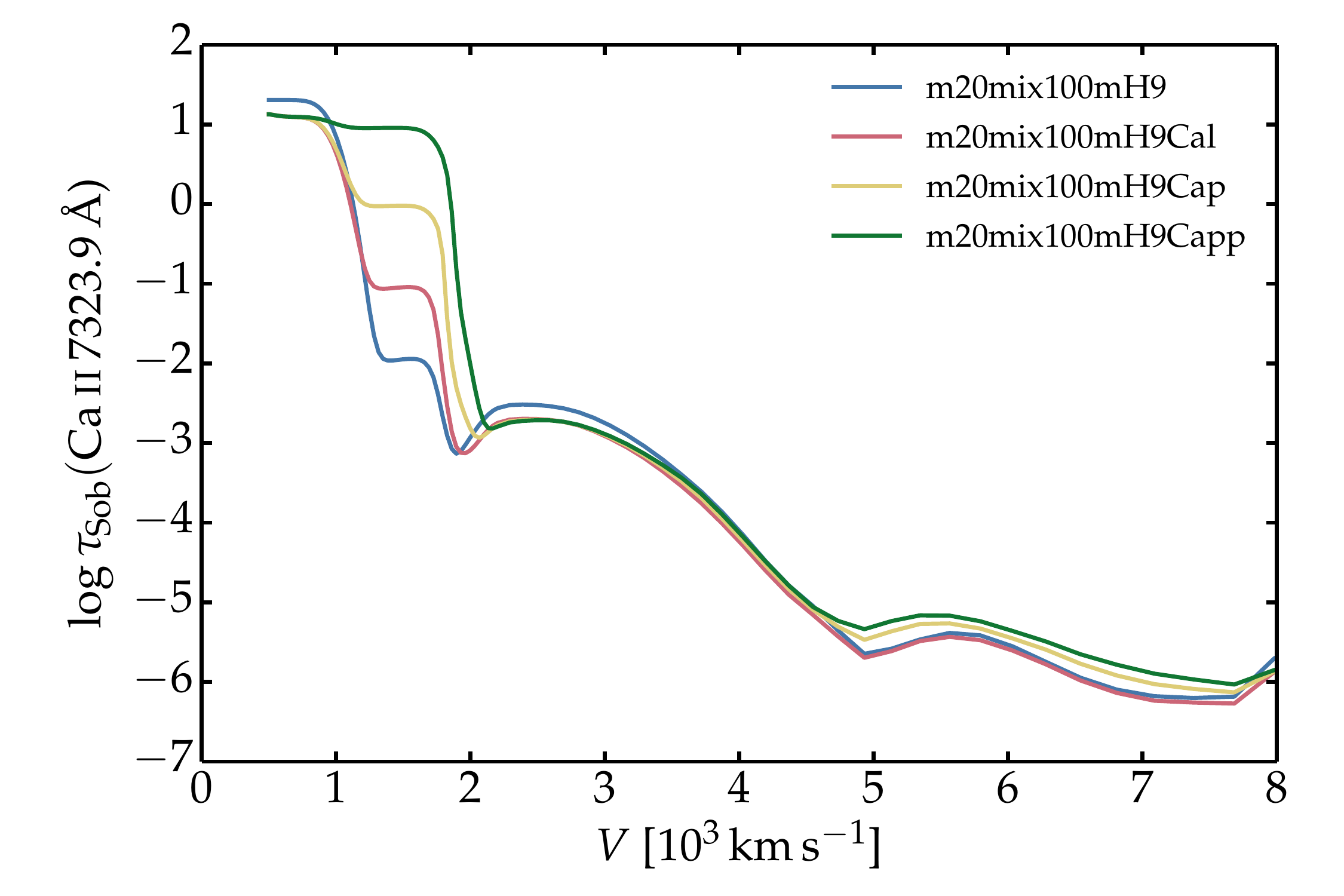}
\caption{Illustration of the Sobolev line optical depth for the Ca\,\two\,7323.9\,\AA\, line for the models shown in Fig.~\ref{fig_Ca_in_Oshell}. 
\label{fig_taul_ca2}
}
\end{figure}

As expected, raising the Ca mass fraction in the O-rich shell has a dramatic effect on the \oidoub\ doublet strength (note that the distribution of the decay power absorbed is the same in all four models since we switch O for Ca but at constant density and mass; see middle panel). The effect is present even for an O/Ca mass fraction ratio in the O-rich shell of 1000. The influence on H$\alpha$ is negligible, which is expected since the H-rich layers were not modified. The Ca\two\ lines are strengthened, but for the higher Ca enhancement, their strength decreases to the benefit of Ca\one\ lines (mostly present in the near infrared; bottom spectral panel of Fig.~\ref{fig_Ca_in_Oshell}). The ionization is still primarily Ca$^+$, but there is a small inflection in the ionization, the electron density, and in the temperature in the O-rich shell. The most likely reason for this saturation and even reduction of the Ca\,\two\ flux is that both doublet components are optically thick (Fig.~\ref{fig_taul_ca2}).

\subsection{Implications and comparison to previous work}

A number of points need to be made at this stage. Rather than being secondary and irrelevant, the Ca mass fraction in the O-rich shell is a critical matter because it affects the O\one\ doublet, which is used for constraining the progenitor mass. In the above experiment, the mass of the O-rich shell is the same in all four ejecta models and yet the O\one\ doublet flux varies by a factor of about five just by tuning the Ca/O ratio in the O-rich shell.

Some of the adopted Ca mass fractions in this experiment are probably too large. However, stellar evolution simulations frequently produce a large Ca mass fraction in the O-rich shell. This occurred in half the \mesa\ simulations we ran for this study. The cause is the merging of the Si-rich shell and the O-rich shell during Si burning, producing a Ca mass fraction of a few 0.001 in the O-rich shell, so that the O/Ca mass fraction ratio drops by a factor of 100 compared to the case of no shell merging (the total Ca mass is however unchanged by this merging). A similar feature was observed by \citet{fransson_chevalier_89} in some of their models (computed with \kepler\ and presented in \citealt{ensman_woosley_88}). \citet{collins_presn_18} report a higher occurrence of merging for the O, Ne, and C burning shells in higher mass progenitors between 16 and 26\,\msun, yielding a large Si mass fraction in a large part of the O-rich shell -- this is the same as what we observe in our \mesa\ simulations. Similarly, \citet{yoshida_presn_19} obtain distinct composition profiles within the Si- and O-rich shells depending on the adopted convective overshoot strength and progenitor mass.

In the absence of such a merging, the Ca mass fraction in the O-rich shell is typically at the original value on the ZAMS (i.e., at the solar metallicity in our models; see section~\ref{sect_caveats} for departures from a solar value), but one can wonder whether this feature would persist in 3D hydrodynamical simulations treating physically the processes of convection and overshoot. Even with strong overshoot from the Si-rich shell, it is not clear that mixing could take place down to the microscopic level given the short time until collapse and explosion (typically a day from the onset of Si burning; see, e.g., \citealt{arnett_book_96} or \citealt{collins_presn_18}) -- the mixing might instead be partial and truncated at some spatial scale.

The origin of Ca\two\ emission is a related issue. In the experiment above, if the Ca mass fraction is large in the O-rich shell (say with a $\sim$\,0.01 mass fraction), most of the power absorbed by the O-rich shell occurs through Ca\two\ lines this process can be mitigated by line optical depth effects). Otherwise, we obtain Ca\two\ emission from the Si-rich shell, the interface between the Si-rich shell and the O-rich shell (where the Ca mass fraction is around or above 0.01), and from the He-rich and H-rich shells.

\citet{li_mccray_ca2_93} and others argue that the Ca\two\ emission arises primarily from the Ca in the H-rich envelope. The main limitation for Ca emission from the Si-rich shell is that the Si-rich shell is of low mass, and hence it absorbs a small fraction of the total decay power. Even if Ca were the sole coolant in that Si-rich shell, the total power that it would radiate would be a small fraction of the total, typically smaller than the power absorbed in the H-rich layers (though this can depend on the mass of the CO core and the level of mixing). More importantly, the main competitor is the CO core mass, which in massive progenitors may absorb most of the decay power (again a function of mixing, etc.). Whether Ca emission comes from the H-rich envelope is not just a matter of Ca being a strong coolant. It depends primarily on where the decay power is absorbed. For a larger He core mass (a higher mass progenitor), for a larger ejecta mass, or for a weaker level of \nifs\ mixing, a larger fraction of the decay power may be absorbed in the core rather than the H-rich layers. In this situation, the H-rich ejecta layers would only reprocess the radiation impinging from below.

The origin of the Ca\two\ emission would be altered in the case of Ca pollution into the O-rich shell prior to core collapse. This would enhance the \caiidoub\ doublet flux at the expense of the \oidoub\ doublet flux. If the process of shell merging had a higher occurrence rate in higher mass progenitors \citep{collins_presn_18}, it would provide a natural explanation for the lack of SNe II with strong \oidoub\ at nebular times, and inferred progenitor masses below 17\,\msun.

\begin{figure*}
\includegraphics[width=\hsize]{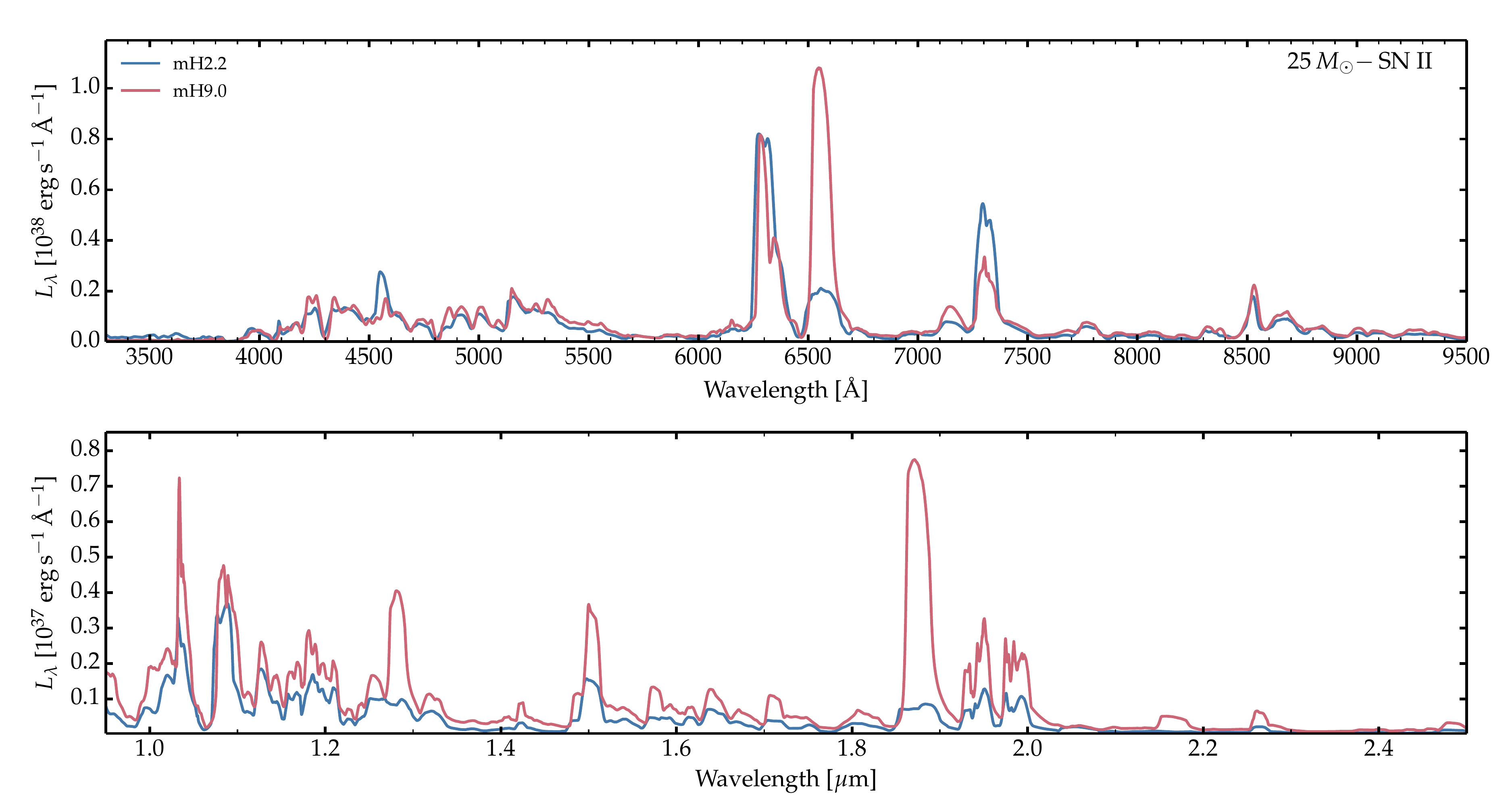}
\includegraphics[width=\hsize]{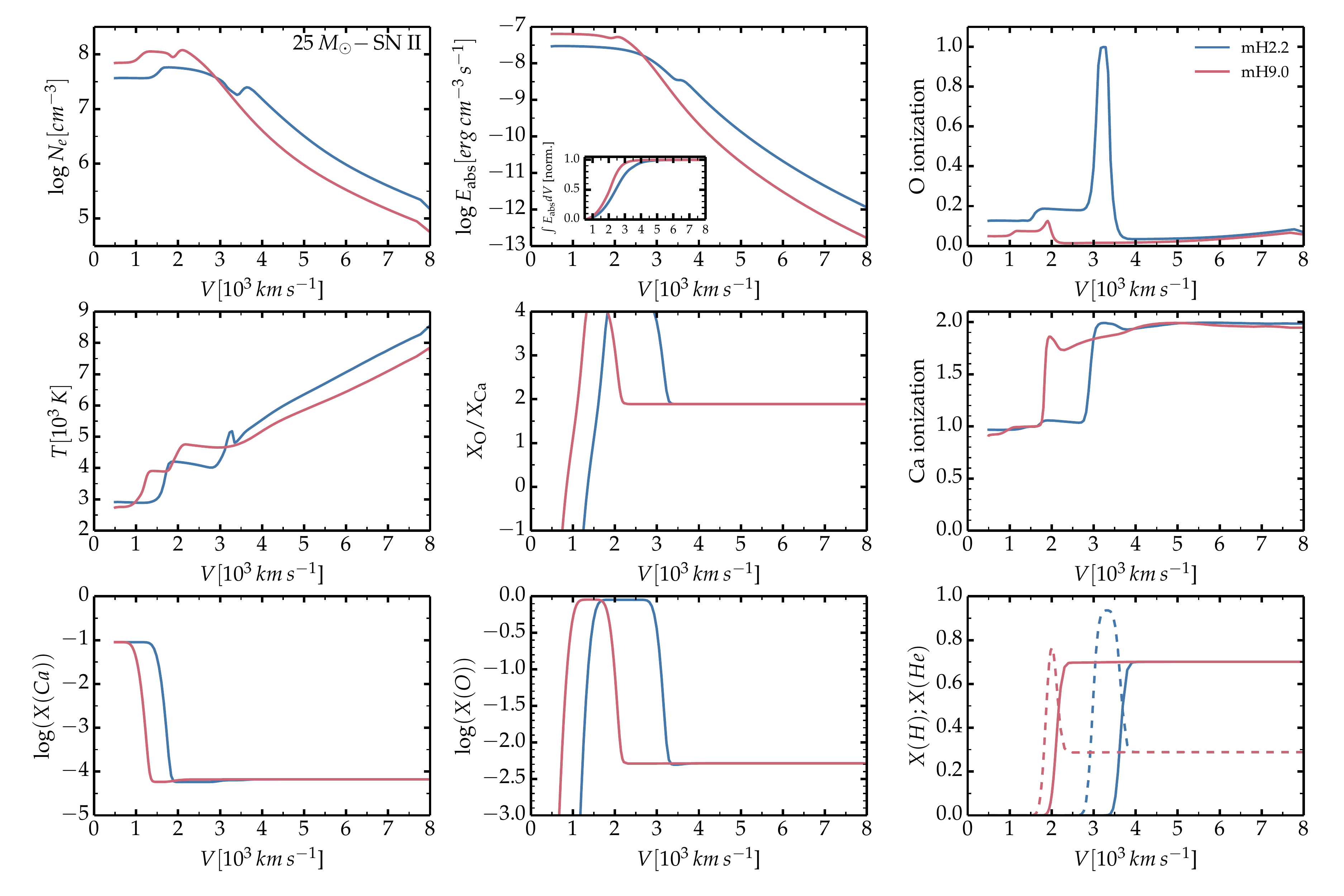}
\caption{Similar to Fig.~\ref{fig_ref_model}, but now showing the influence of the progenitor H-rich envelope mass on the synthetic optical and near-infrared spectra at 300\,d. The models shown are m25mix100vni and m25mix100vnimH9 (strong mixing of \nifs\ but weak mixing of other species) and correspond to a 25\,\msun\ progenitor star. There is a greater fraction of the total decay power absorbed in the H-rich layers of the lighter model m25mix100vni, boosting the \caiidoub\ doublet emission, but this has an adverse effect on H$\alpha$, which pushes its formation to the outer lower density regions. These differences reflect in part the different distribution of decay-power absorbed since the two ejecta have a different velocity structure -- mixing should also differ in nature for  ejecta with  a similar kinetic energy but a different mass (e.g., as in SNe II and Ibc).
}
\label{fig_m25_mhenv}
\end{figure*}

\section{Influence of the H-rich envelope mass}
\label{sect_henv}

In our toy setup, we either force the ejecta mass to be 10\,\msun\ (and we adjust the H-rich envelope mass) or we force the H-rich envelope mass to be 9\,\msun\ (and we adjust the ejecta mass to be the sum of the H-rich envelope mass plus the He-core mass). The latter might be a good description of single stars initially in the mass range between 12 and about 20\,\msun\ and the latter more suitable to higher mass stars but still dying as RSGs \citep{whw02,d19_sn2p}. Since we assume the same ejecta kinetic energy, variations in total mass or envelope structure will impact the distribution of elements in velocity space.

Figure~\ref{fig_m25_mhenv} illustrates the impact of the H-rich envelope mass of the progenitor (or the total mass of the ejecta) for a 25\,\msun\ progenitor model (this defines the core properties). The low-envelope mass model is m25mix100vni (total ejecta mass is 9.9\,\msun) and the high-envelope mass model is m25mix100vnimH9 (total ejecta mass is 16.7\,\msun) in our nomenclature. With the higher H-rich envelope mass, a greater fraction of the decay power emitted is absorbed (90\% compared to 67\%). This implies a general flux offset. A sizable fraction of this extra power is radiated in H\one\ lines. Furthermore, for a greater ejecta mass and the same ejecta kinetic energy, the metal-rich regions are located at smaller velocities. This causes a small reduction in the width of lines associated with O\one\ and Mg\one\ (lines forming in the extended H-rich layers are less affected). That effect is exacerbated by the greater ejecta density. This, in turn, enhances the local deposition of the decay power in the slower denser layers that contain more \nifs\ initially (the $\gamma$-ray mean free path is shorter). Hence, the progenitor H-rich envelope not only affects the type II-P SN radiation during the photospheric phase (by modifying its length and brightness), it can also influence the nebular phase properties.

\section{Influence of \nifs\ mixing}
\label{sect_ni_mix}

\begin{figure*}
\includegraphics[width=\hsize]{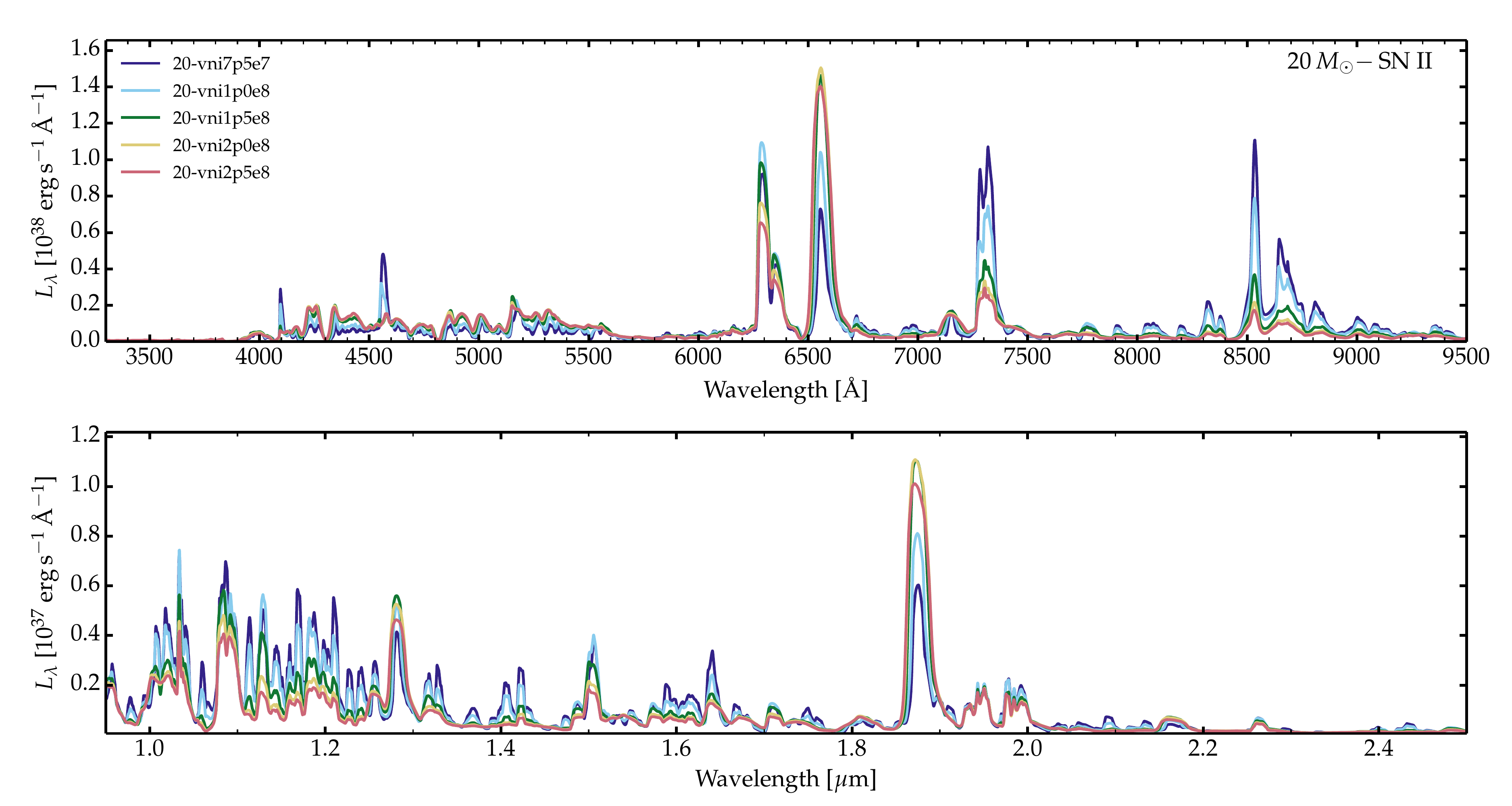}
\includegraphics[width=\hsize]{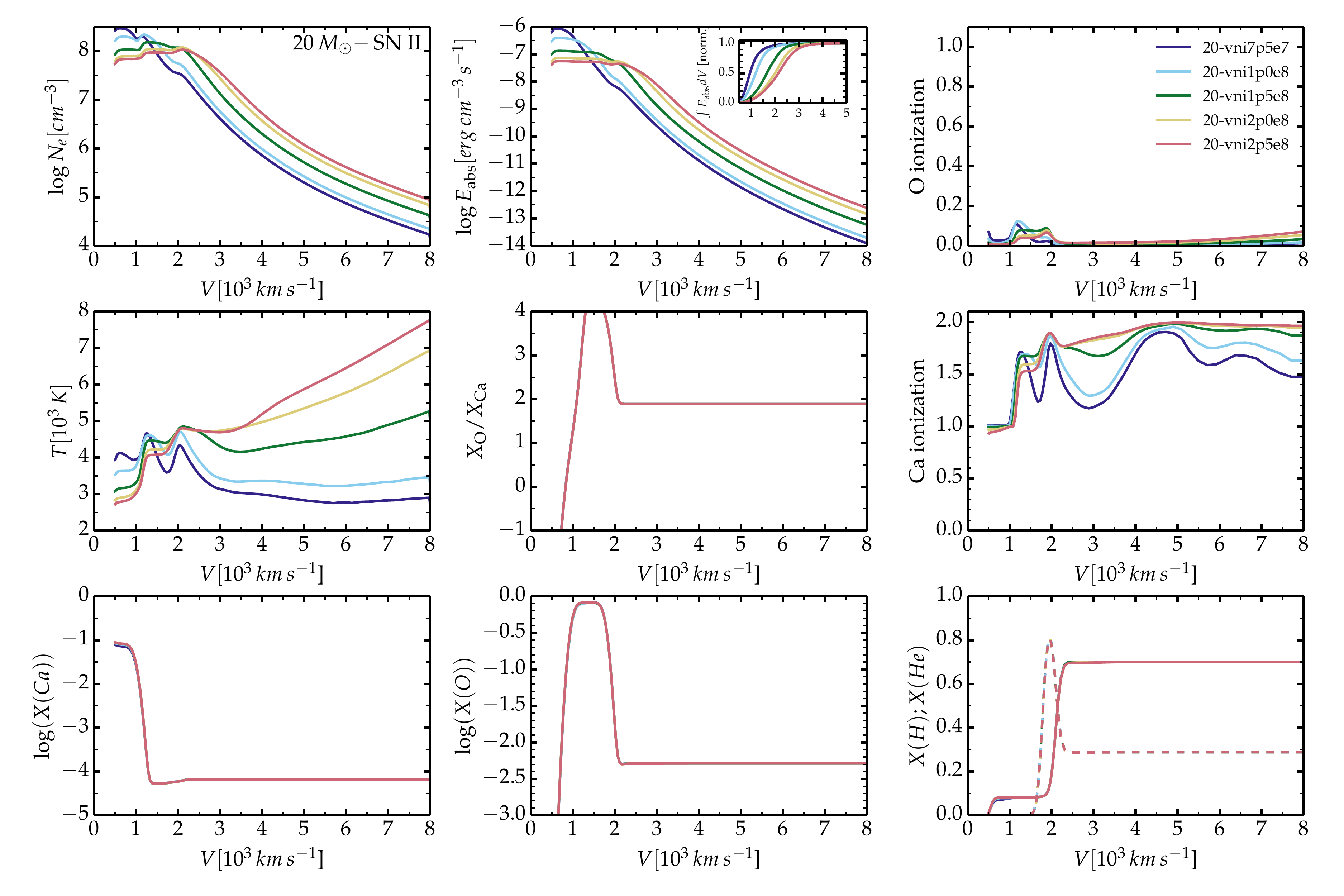}
\caption{Similar to Fig.~\ref{fig_ref_model}, but now showing the influence of the \nifs\ mixing on the SN radiation and gas properties at 300\,d. The ejecta arises from a 20\,\msun\ progenitor, weak mixing for all species, imposed H-rich envelope of 9\,\msun, and a \nifs\ mass of 0.08\,\msun. An additional and variable amount of \nifs\ mixing is applied increasing $V_{\rm Ni}$ from 750 (model 20-vni7p5e7) to 2500\,\kms\ (model 20-vni2p5e8). With enhanced \nifs\ mixing, more decay power is absorbed in the outer ejecta, causing a strengthening of H$\alpha$ but a weakening of \caiidoub\ because of the boost to the Ca ionization (the ionization shift also affects Fe\one\,--\two\ and Mg\one\ lines in the optical and the near-infrared). [See section~\ref{sect_ni_mix} for discussion.]
\label{fig_Hincore_var_nifs_mix}
}
\end{figure*}

In this section, we explore the influence of \nifs\ mixing on the SN radiation properties. We use the core properties of a 20\,\msun\ progenitor model and assume a 9\,\msun\ H-rich envelope. Although not essential, we set in all models a 10\% H mass fraction throughout the core layers (Si, O, and He-rich shells) to mimic the inward mixing of H-rich material at low velocity (in real explosions, outward mixing of core material occurs simultaneously with inward mixing of envelope material). A weak mixing is applied to all species, and a subsequent additional mixing is applied to \nifs\ with $V_{\rm Ni}$ covering from 750 to 2500\,\kms\ in four increments (five models). $\Delta V_{\rm Ni}$ is set to $V_{\rm Ni}/2.5$ (see Eq.~\ref{eq_vni}). Results are shown in Fig.~\ref{fig_Hincore_var_nifs_mix}.

The ejecta model with weak mixing completely traps the decay power while the ejecta model with the highest mixing of \nifs\ lets 10\% of the decay power escape at 300\,d. This effect is difficult to discern from observations since about 30\% of the total flux falls outside of the optical range at that time (and is thus rarely measured). Besides facilitating escape, enhanced mixing favors the more extended deposition of decay power. In the weaker mixing model, nearly 70\% of the decay power is deposited below 1000\,\kms\ and thus within the Si-rich shell (with a small portion in the base layers of the O-rich shell). Consequently, a significant fraction of the SN radiation arises from the Si-rich and O-rich layers, as evidenced by the narrow Si, Ca, and Fe line emission. Similarly, the faster moving H-rich layers capture little of this decay power so that H\one\ lines are weak and quite narrow, and there is a lack of broad Fe lines.

As the \nifs\ mixing is enhanced, it first benefits the O-rich shell which produces a stronger \oidoub\ doublet. As the \nifs\ mixing is enhanced to its maximum, the \oidoub\ doublet strength drops again as a greater fraction of the decay power is now absorbed in the H-rich shell. Consequently, H\one\ and Fe\one\,--\two\ lines strengthen and broaden.

Interestingly, the \caiidoub\ doublet weakens and does not broaden, which implies that the greater power absorbed by the H-rich shell is not radiated by \caiidoub. The enhanced \nifs\ tends to raise the temperature and ionization (see the electron density) with increasing ejecta velocity. However, the influence on the ionization of a given species is more complicated. In the present set of simulations, the O ionization remains unchanged, but the Ca ionization is enhanced, nearly to Ca$^{2+}$ everywhere beyond 1000\,\kms\ for the model with the highest mixing.\footnote{The ionization potential of O\one\ is 13.6\,eV, and it takes another 35.1\,eV to ionize O$^+$. The ionization potentials of Ca\one\ and Ca\two\ are low with only 6.1 and 11.9\,eV, while it takes 50.9\,eV to ionize Ca$^{2+}$. These ionization energies and the low temperature of type II SN ejecta favor the formation of lines from O\one\ and Ca\two\ in SN II nebular spectra.} This increase in Ca ionization quenches the \caiidoub\ doublet emission. This effect was previously seen in a study of Type II-P SN  by \cite{1988Ap.....29..449C} who attribute it to the influence of Ly$\alpha$ photons. The change in electron density is largely driven by the change in H ionization.

Variations in \nifs\ mixing, and their impact on ejecta ionization, may contribute in part to the diversity of nebular-phase spectra  of type II SNe (see, for example, the discussion in \citealt{yuan_13ej_16}).

By simply varying the \nifs\ profile, this experiment shows that spectral lines at the nebular epoch can be considerably tuned in strength and width for the same ejecta composition, mass, and kinematics. This is well understood but it is a concern since \nifs\ mixing is a complicated process, with \nifs\ probably having a unique distribution in each SN. This is most likely tied to additional dependencies on progenitor mass and rotation, He-core mass, and explosion energy \citep{wongwathanarat_15_3d,sukhbold_ccsn_16}. It also shows that metal line emission (in particular of Ca and Fe in our limited composition mixture) can arise primarily from the Si-rich shell or from the H-rich shell depending on where the decay power is absorbed. The conclusion of \citet{li_mccray_ca2_93} that \caiidoub\ arises from the H-rich material (which they infer from their analysis of SN\,1987A) only holds if \nifs\ mixing is strong so that most of the decay power gets absorbed by the H-rich material at large velocities (an alternative is that the H-rich material is mixed inwards closer to \nifs\ and thus absorbs decay power in the absence of strong \nifs\ mixing). A further issue is that Ca over-ionization can completely inhibit \caiidoub\ emission from the H-rich layers, no matter how much decay power is absorbed there.

\begin{figure*}
\includegraphics[width=\hsize]{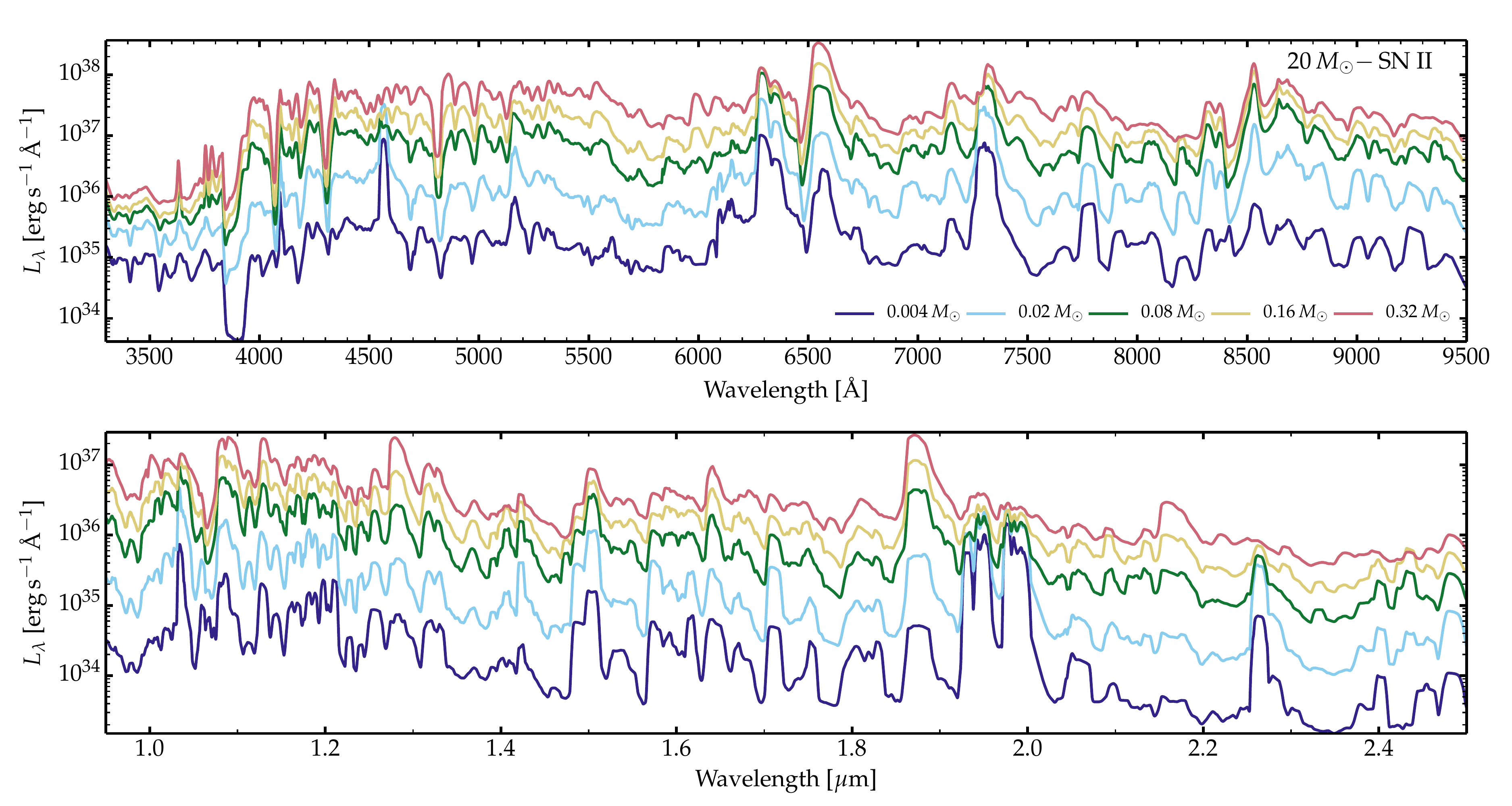}
\includegraphics[width=\hsize]{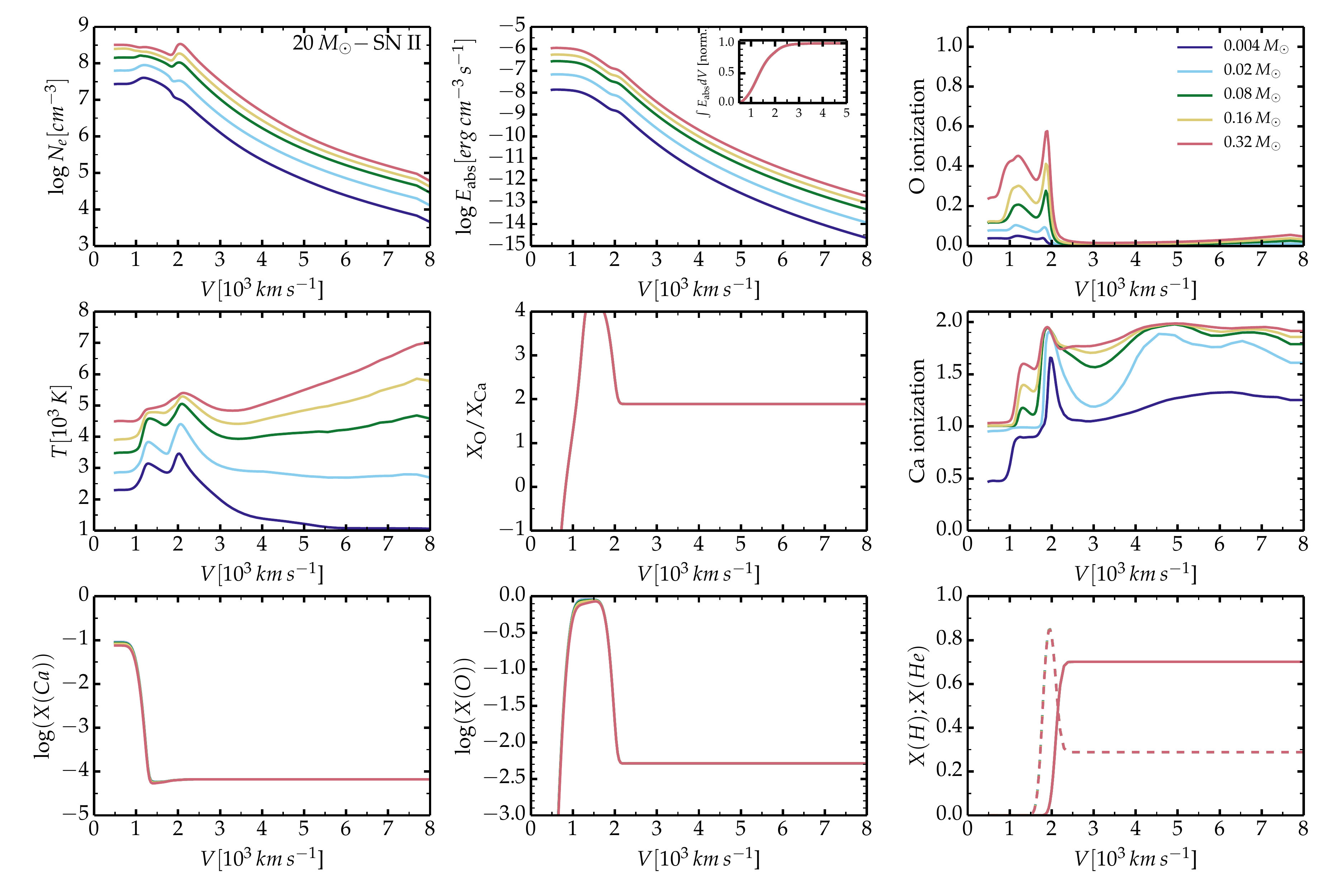}
\caption{Similar to Fig.~\ref{fig_ref_model}, but now showing the influence of the \nifs\ mass on the SN radiation (for better visibility, a logarithmic scale is used for the luminosity). The ejecta arises from a 20\,\msun\ progenitor, weak mixing for all species, no additional mixing of \nifs, imposed H-rich envelope of 9\,\msun, and is characterized by a \nifs\ mass of 0.004, 0.02, 0.08, 0.16, and 0.32\,\msun. A greater \nifs\ mass boosts the electron density and species ionization, inhibits forbidden line emission (in favor instead, for example, of recombination lines), and  strengthens optical depth effects and continuum processes (primarily in association with electron scattering). [See section~\ref{sect_var_mni} for discussion.]
\label{fig_m20_var_mni}
}
\end{figure*}

\section{Dependencies on \nifs\ mass}
\label{sect_var_mni}

   We now discuss the influence of the \nifs\ mass on the resulting optical and near-infrared spectra as well as on the ejecta properties. For this exploration, we start off with model  m20mix100mH9 (20\,\msun\ progenitor, weak mixing for all species, no additional mixing of \nifs, imposed H-rich envelope of 9\,\msun), characterized by a \nifs\ mass of 0.08\,\msun\ and compute additional models in which the \nifs\ mass is reset to 0.004, 0.02, 0.16, and 0.32\,\msun. Results are shown in Fig.~\ref{fig_m20_var_mni}.

All five ejecta models achieve complete $\gamma$-ray trapping so there is a linear scaling between the \nifs\ mass and the bolometric luminosity. We also find that all five models radiate the same fraction of the total power within the optical range (about 75\%). However, the optical spectra vary considerably with \nifs\ mass. There are various reasons for this.

For higher \nifs\ mass, the ejecta ionization is greater, yielding amongst other things an increased electron density. Going here from  0.004 to 0.32\,\msun, which is certainly not a small range, the total radial electron-scattering optical depth to the core increases by more than a factor of ten (it grows from 0.08 to 1.1). One should recall that typical type II SN ejecta turn optically thin within a few months because of recombination. If somehow this recombination can be prevented, as can be done for example by supplying power from a magnetar \citep{d18_iptf14hls}, the ejecta can stay optically thick for two years after explosion.

Here, the effect is not as strong but shows the same trend. Consequently, the model with the lowest \nifs\ mass is very optically thin (total electron-scattering optical depth of 0.08 and total flux-mean optical depth of 0.8), while the model with the highest \nifs\ mass is not optically thin (total electron-scattering optical depth of 1.0 and total flux-mean optical depth of 2.2), even at 300\,d. In the latter, optical depth effects and collision processes are not negligible. The greater electron density places many forbidden lines above their critical density, which will quench their emission (i.e., upper levels can more frequently de-excite through collisions).  A larger \nifs\ mass also leads to a greater ionization of numerous species such as O and Ca, and to a higher temperature which facilitates populating excited states.

Summarizing, the nebular spectrum in the model with the largest \nifs\ mass shows a greater flux (built from the combination of many overlapping weak and broad emission lines), and a stronger contribution from the Fe\two\ line forest (many of which are permitted lines). For higher \nifs\ mass, the forbidden lines of O\one\ and Ca\two\ are relatively weaker. The \oidoub\ doublet flux represents 30\% (5\%) of the optical flux in the model with 0.004\,\msun\ (0.32\,\msun) of \nifs. Similarly, and in the same order, the \caiidoub\ doublet represents 20\% (6\%) of the optical flux. In addition to the greater electron density, it is also the shift to a higher element ionization that quenches these lines (see O and Ca ionization in Fig.~\ref{fig_m20_var_mni}).

This exploration shows that the \nifs\ mass acts in a very complicated way on the ejecta properties and yields a very non-linear evolution of spectral properties. This result is a warning against using an event like SN\,1987A as a reference to which linearly scale various quantities.

\begin{figure*}
\includegraphics[width=\hsize]{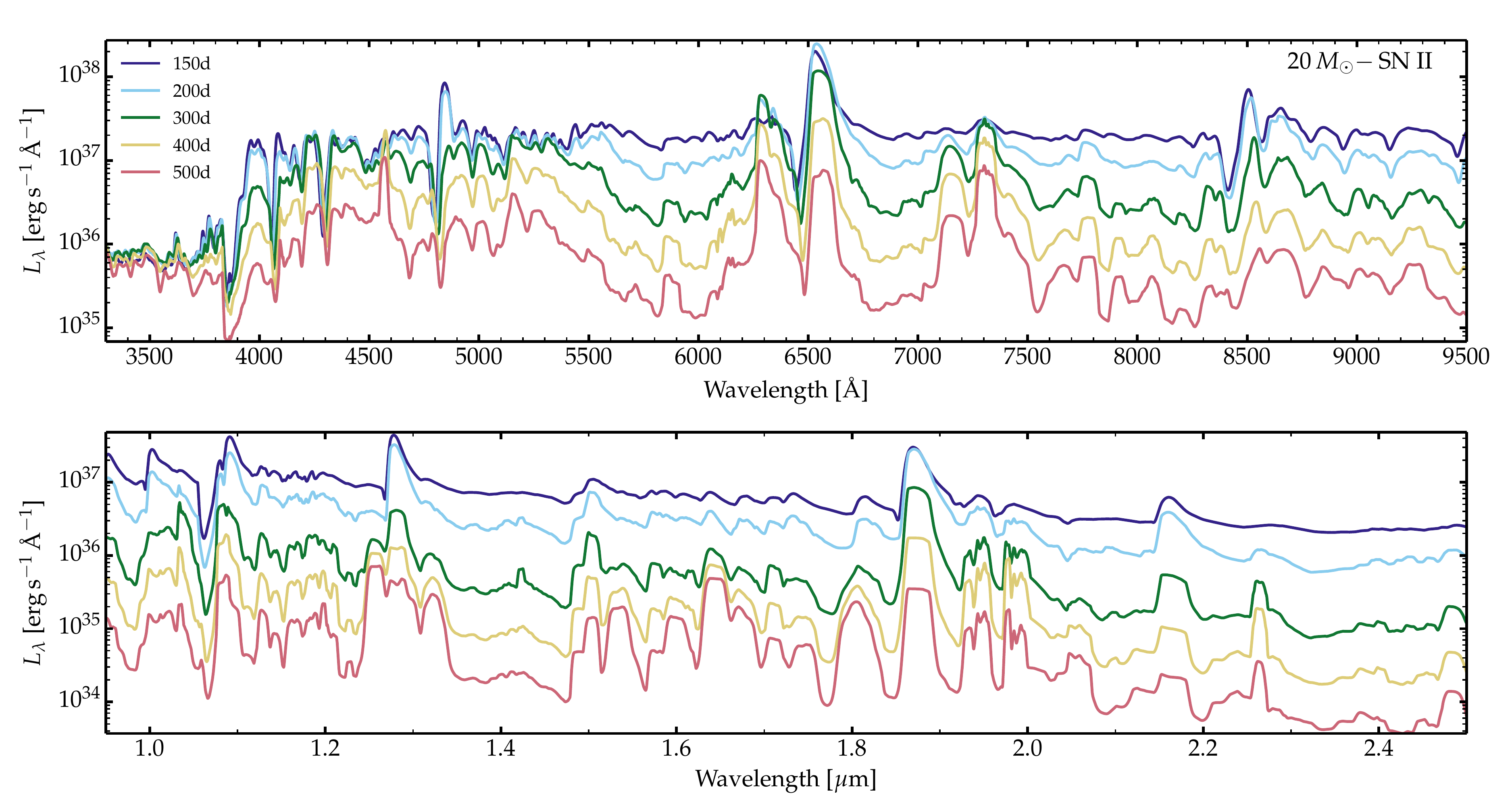}
\caption{Similar to Fig.~\ref{fig_ref_model}, but now showing the evolution of the optical and near-infrared properties of model m20mix100vnimH9 from 150 to 500\,d after explosion. With increasing time, the conditions become increasingly "nebular", with a relative strengthening of forbidden line emission, and the progressive reduction of optical depth effects (for example with the disappearance of P-Cygni profiles). This evolution is similar to reducing the \nifs\ mass at a given SN age (see Fig.~\ref{fig_m20_var_mni}).
\label{fig_m20_evol_time}
}
\end{figure*}

\begin{figure}
\includegraphics[width=\hsize]{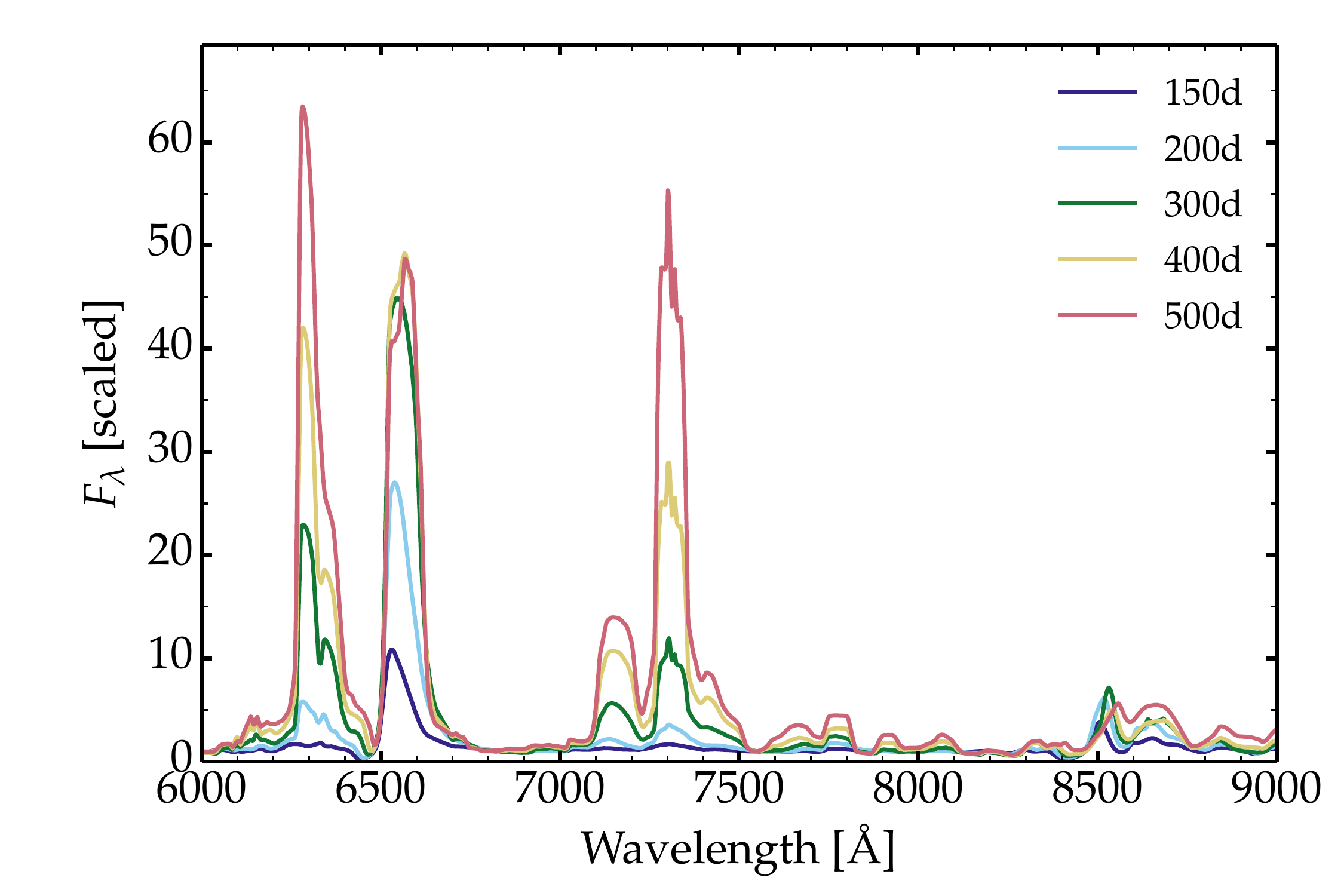}
\caption{As for Fig.~\ref{fig_m20_evol_time}, but using a linear scale for the region covering the \oidoub\ doublet and the Ca\two\ near-infrared triplet.
\label{fig_m20_evol_time_o1_ca2}
}
\end{figure}

\section{Evolution during the nebular phase}
\label{sect_neb_evol}

   Figure~\ref{fig_m20_evol_time} illustrates the evolution of the model luminosity through the optical and near-infrared ranges from 150 to 500\,d for the type II SN model m20mix100vnimh9. The continuous drop in luminosity in both spectral domains results from the drop by a factor of about 23 in the decay power emitted and to a lesser extent the increasing $\gamma$-ray escape with time: 98\% (66\%) of the decay power is trapped at 150\,d (500\,d) after explosion. The simultaneous drop in density (as $1/t^3$) and ionization level (as material cools and recombines) leads to a drop in electron density. Since free electrons are the dominant source of continuum opacity in type II SNe ejecta, this leads to a drop in electron-scattering ejecta optical depth, here from a value of 2.8 at 150\,d to only 0.073 at 500\,d (a value of $\sim$0.3 would result at constant ionization; the corresponding drop in flux-mean optical depth is from 3.3 at 150\,d to 0.37 at 500\,d). This change in electron density has numerous implications (see also discussion in section~\ref{sect_var_mni}).  As the electron density drops below their critical densities, forbidden lines strengthen relative to other lines (since the decay power absorbed in the ejecta decreases with time, the strength of all lines tends to decrease). For O\one, this is also facilitated by the reduction in ionization which populates the ground state of neutral oxygen.

   More subtle properties can be seen from Fig.~\ref{fig_m20_evol_time_o1_ca2}, which focuses on the optical range between the \oidoub\ doublet and the Ca\two\ near-infrared triplet. The individual components of the \oidoub\ doublet have nearly equal strength at 150\,d, which implies that the lines are optically thick, while at 500\,d their relative strength is about three (i.e.,  equal to the ratio of their radiative de-excitation rates $A_{ul}$; see Table~\ref{tab_o1_ca2}). The changing flux ratio of the doublet components is discussed in detail by \citet{li_mccray_o1_92}. Time also has a profound impact on the relative strengths of Ca\two\ lines. The \caiidoub\ doublet strengthens considerably with time relative to the Ca\two\ near-infrared triplet, a consequence of the drop in electron density (see also \citealt{li_mccray_ca2_93}). Although a limitation of assuming spherical geometry, numerous forbidden lines at 500\,d have a  flat-top profile, indicative of emission from a hollow sphere in velocity space (the sphere is in fact a shell, thus bounded between two large velocities).\footnote{For a Hubble flow the escape probability is the same in all directions, independent of whether the line is optically thick or thin. In this context, any line has a flat-topped profile if it forms from a hollow shell. At earlier times the continuum opacity, and blending, will modify the profile shape.}

\begin{figure*}
\includegraphics[width=\hsize]{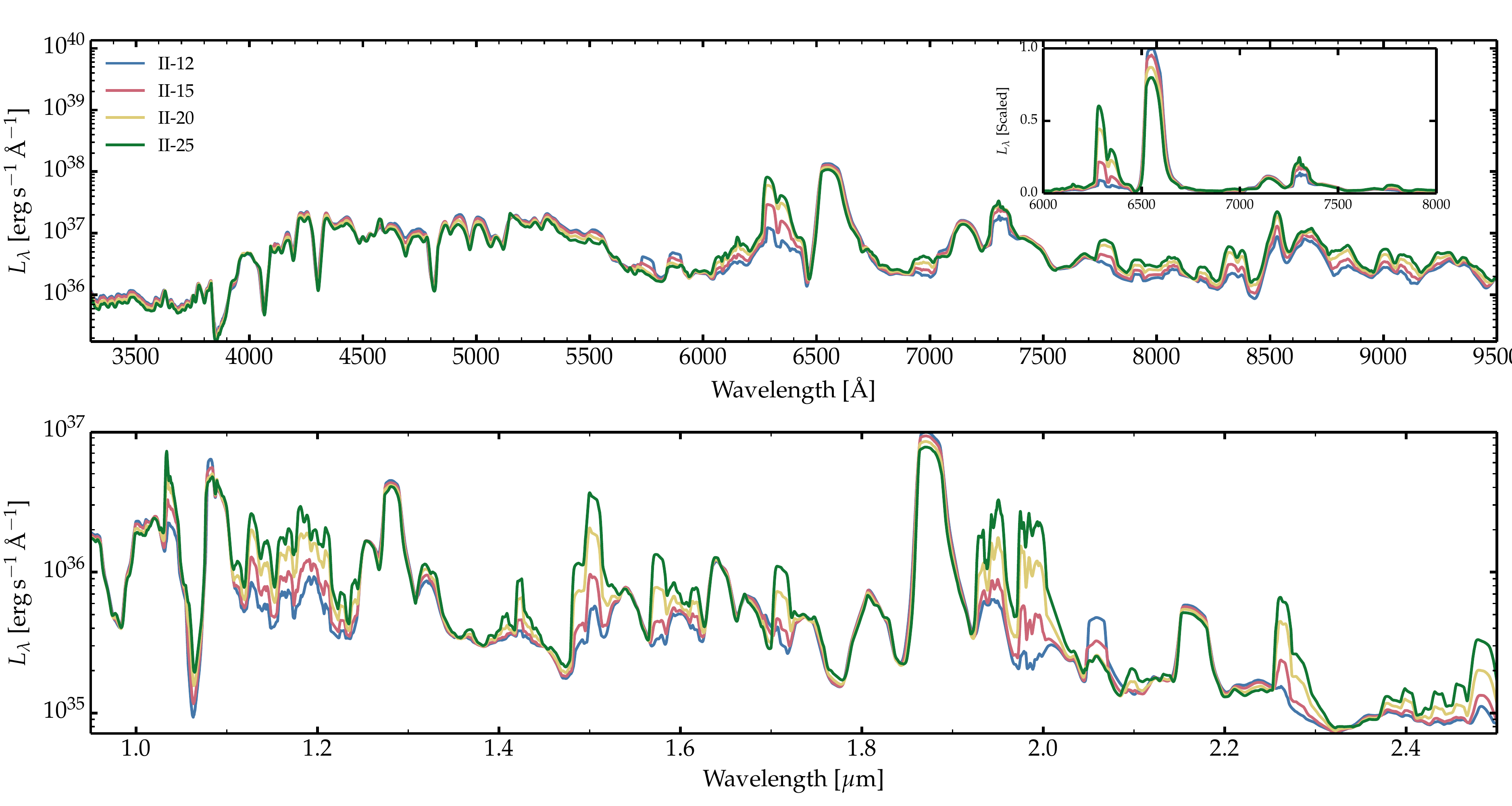}
\caption{Similar to Fig.~\ref{fig_ref_model}, but now showing the optical and near-infrared spectra for type II SN models associated with different ZAMS masses from 12 to 25\,\msun. These models are all characterized by the same H-rich envelope mass of 9\,\msun\ in the progenitor, by weak mixing of non-IGE species (100\,\kms), and by strong mixing of \nifs\ and daughter isotopes. All models in this set have an ejecta kinetic energy of 10$^{51}$\,erg and a \nifs\ mass of 0.08\,\msun.}
\label{fig_toy_II}
\end{figure*}

\section{Signatures of main-sequence mass for type II SNe and uncertainties}
\label{sect_prog_mass}

Figure~\ref{fig_toy_II} shows the model luminosity through the optical and near-infrared ranges for the type II SN models arising from ZAMS masses of 12, 15, 20, and 25\,\msun. The model names are, in this order, m12mix100vnimH9, m15mix100vnimH9, m20mix100vnimH9, and m25mix100vnimH9, and all are characterized by weak mixing for all species except for the \nifs\ and daughter products, which are strongly mixed.

As is well known, for the same decay absorbed in the metal-rich layers, the strength of the \oidoub\ doublet increases with ZAMS mass (see for example \citealt{fransson_chevalier_89} or \citealt{jerkstrand_04et_12}). We obtain this result because we did not force strong mixing of elements (preventing the pollution of Ca from the Si-rich layers with the O-rich material; see section~\ref{sect_o_over_ca}). The systematic increase of  the \oidoub\ doublet luminosity with ZAMS mass arises from the increase in decay power absorbed by the O-rich layers relative to other layers -- the O-rich layers absorb a greater share of the total power available. This occurs because of the increase in CO core mass with ZAMS mass. This increase is exacerbated by the reduced O ionization with increasing progenitor mass, so that O is essentially neutral in the O-shell (in fact throughout the ejecta) in the 25\,\msun\ model. Interestingly, probably because of the strong \nifs\ mixing in this model set, Ca is twice ionized throughout the H-rich layers, which quenches Ca\two\ line emission from the outer ejecta. But because of strong \nifs\ mixing, decay power is preferentially absorbed outside of the Si-rich layers, which quenches Ca\two\ line emission in the inner ejecta. Consequently, Ca\two\ lines are weak in this model set.

One lesson to draw from this and previous sections is that the \oidoub\ doublet emission is a much more robust indicator than the \caiidoub\ doublet because \oidoub\ predominantly arises from the O-rich shell and because the O yield strongly scales with ZAMS mass. Ca is most abundant in the low-mass Si-rich shell (whether in the progenitor star or after explosive nucleosynthesis), which absorbs little power, and the associated Ca\two\ emission is often spread over multiple shells/regions in the ejecta. The spreading depends on the \nifs\ mixing, which is not well constrained from observations. Finally, the mass of the explosively produced Si-rich shell is sensitive to the physics and the properties of the explosion (see, for example, \citealt{WH07}).

\section{Comparison to observations}
\label{sect_obs}

\begin{figure*}
\begin{center}
\includegraphics[width=0.95\hsize]{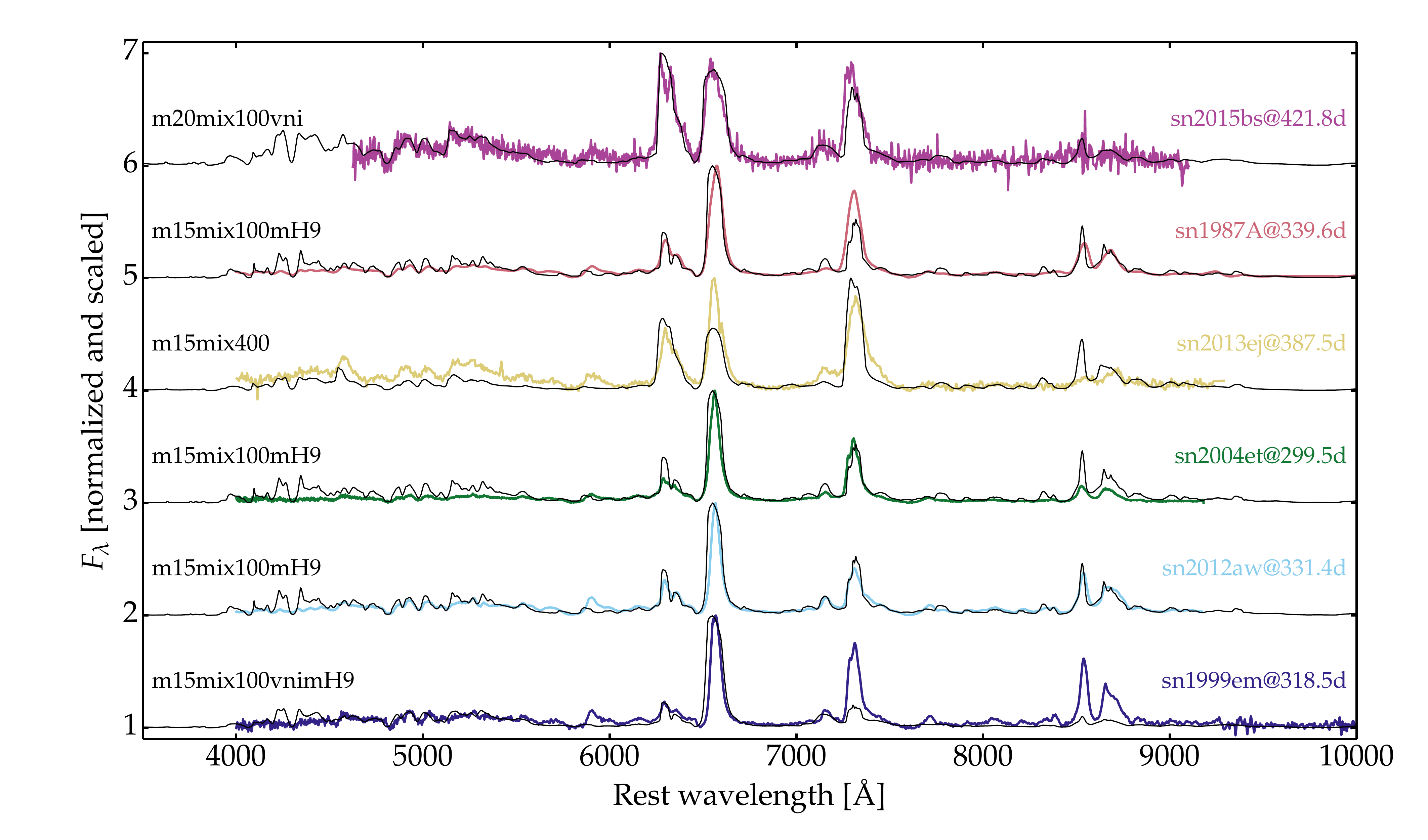}
\end{center}
\caption{Comparison of optical spectra for a sample of observations and a selection of models presented in this work -- we make no attempt at obtaining best fits. Apart from SN\,2015bs, which stands apart, most of these well-observed SNe II show similar spectral properties at $300-400$\,d after explosion. A variety of models is used to show the offsets caused by certain assumptions (for example,  the weak \caiidoub\ that follows from Ca over-ionization caused by strong \nifs\ mixing in model m15mix100vnimH9).  Model m15mix100mH9, used for SN\,2012aw, would provide a better match to the observations of SN\,1999em.
\label{fig_comp_to_obs}
}
\end{figure*}

Having produced a grid of models, we can now compare our synthetic spectra to a few well observed type II SNe in the nebular phase. We have selected a few events with standard photometric and spectroscopic properties (brightness during the photospheric phase, ejecta mass and kinetic energy, \nifs\ mass) so that they likely reside in a similar parameter space to our models. Our selection is detailed in Table~\ref{tab_obs} and includes SNe 1987A, 1999em, 2012aw, 2004et, 2013ej, and 2015bs. Figure~\ref{fig_comp_to_obs} shows the comparison between observations (spanning a SN age of 300 to 420\,d) to the model spectra (all at 300\,d after explosion).

All type II SNe in our sample can be fitted with one or several of our 15\msun\ models except for SN\,2015bs, which requires both a more massive progenitor and peculiar properties. The models with weak mixing (mix100) yield satisfactory matches to observations, for example for SN\,2012aw. The width of the \oidoub\ doublet is well matched which suggests that the O-rich material was not extensively mixed. In the corresponding model m15mix100mH9, there is also weak \nifs\ mixing so that the bulk of the decay power is absorbed in the metal-rich core. The presence of a massive H-rich envelope in the progenitor implies that the H-rich material extends down to low velocities (composition stratification), irrespective of the weak mixing. This model also fits reasonably well SN\,1987A, although some lines are too narrow or lack extended wings (the emission from the H-rich envelope is underestimated, which arises from the insufficient \nifs\ mixing). Because there is little mixing of H-rich material down to low velocities, the H$\alpha$ profile generally lacks emission at low velocity (the line appears less triangular than observed).

In model m15mix100vnimH9 (used for the comparison to SN\,1999em), the enhanced mixing produces broader lines overall, but the smaller power absorbed in the inner metal rich regions leads to weaker Ca\two\ emission from the inner ejecta (the narrow part of the line) while the greater power absorbed in the H-rich layers boosts the Ca ionization and quenches Ca\two\ emission. Weak Ca\two\ emission is rare but has been observed in SN\,2012ec \citep{jerkstrand_ni_15}. Because of the similar spectral appearance of SN\,1999em and SN\,2012aw, model m15mix100mH9 would yield a satisfactory match to SN\,1999em (the goal here is not to reproduce observations but to identify trends and processes).

The broad O\one\ and Ca\two\ lines in SN\,2013ej are hard to match with our grid of models. We find that the models with strong mixing of all species fare best. Since such a mixing may not occur in Nature, this match may be artificial. It is remarkable to see that despite the very crude setup for our ejecta, most models appear rather close to observations. One intriguing example is SN\,2015bs for which model m20mix100vni reproduces quite closely the large widths of \oidoub, \caiidoub, and H$\alpha$. The reason behind this is the strong \nifs\ mixing and the low H-rich envelope mass, allowing the abundant metal-rich material to absorb a large fraction of the decay power and expand much faster than the other SNe in the sample. In addition, the low level of mixing ensures that hydrogen is located at large velocities. Irrespective of these mixing properties, only a high mass progenitor (here  a 20\,\msun\ model is used) can match these properties, as proposed by \citet{anderson_15bs_18}.

As mentioned earlier, the idea was not to provide a match to all existing type II SN spectra in the nebular phase since we craft our models. The dependencies we find are however indicative of the ejecta properties and processes that control the appearance of Type II SN spectra. Our future work will use physical models for both the progenitor and the explosion physics (as previously done, for example, in \citealt{d13_sn2p}), as well as a more suitable technique for treating the process of mixing with greater physical consistency.

   \begin{table}
\caption{Characteristics of the observed Type II SNe used in this paper, including the inferred time of explosion, the redshift, the distance, the reddening, and the reference from whence these quantities and observational data were taken.
\label{tab_obs}
}
\begin{center}
\begin{tabular}{l@{\hspace{3mm}}c@{\hspace{3mm}}c@{\hspace{3mm}}
c@{\hspace{3mm}}c@{\hspace{3mm}}c@{\hspace{3mm}}
}
\hline
                             &   $t_{\rm expl}$    &   $z$     &     $d$       &       $E(B-V)$       &     Ref.   \\
                             &         [MJD]                 &              &    [Mpc]    &           [mag]        &                \\
\hline

SN\,1987A &      46849.82  &    0.00088 & 0.05    &  0.15  &  a \\
SN\,1999em   &   51474.3 & 0.0024 & 11.5 & 0.1 &  b \\
SN\,2012aw    &  56002.6  & 0.0026 & 9.9 & 0.074 &  c \\
SN\,2004et   &  53270.5  &  0.0009 & 7.73 & 0.3 &  d \\
SN\,2013ej    &  56497.5  & 0.0022 & 10.2  & 0.06 &  e \\
SN\,2015bs &      56920.60  &   0.027   &     120.23  &  0.04 &   f  \\
\hline
\end{tabular}
\end{center}
\flushleft
{\bf Notes:} The references used are:
a: \citet{phillips_87A_88,phillips_87A_90};
b: \citet{leonard_99em} and \citet{DH06};
c:~\citet{dallora_12aw_14};
d: \citet{sahu_04et_06} -- we use a lower reddening $E(B-V)$ of 0.3\,mag; We adopt the distance of \citet{vandyk_17eaw_19}.
e: \citet{yuan_13ej_16};
f: \citet{anderson_15bs_18}.
\end{table}

\section{Conclusions}
\label{sect_conc}

We have presented a grid of non-LTE steady-state calculations for type II SNe in order to examine the influence of various parameters on nebular spectra. Results from stellar evolution calculations were used to craft ejecta models. Critical properties at core collapse of stars in the mass range $12-25$\,\msun\ include the monotonically increasing He-core mass with ZAMS mass or the typical abundance ratios within the main shells (i.e., the Si-rich, O-rich, He-rich, and H-rich shells). Variations were introduced in ejecta properties that may show an erratic behavior from SN to SN. This concerns especially the properties of mixing, the abundance of \nifs, or the mass of the H-rich envelope at the time of explosion. With this flexible approach, we examined the dependencies of nebular-phase spectra of type II SNe on variations in ejecta properties but keeping control of the robust differences seen in massive star evolution models.

We first identified a sensitivity of our results to line overlap which is enhanced when we adopt a Doppler width of 50\,\kms\  for all species (which is the standard procedure in our SN \cmfgen\ calculations).  With this Doppler width some Fe\two\ and O\one\ lines overlap with Ly\,$\alpha$ and Ly\,$\beta$. Reducing the Doppler width to 2\,\kms\ reduces the overlap, and leads to significant changes in the resulting spectrum, and in particular to an increase in the H$\alpha$ line strength. Other lines are also affected but their behavior is model-dependent.  Most simulations presented in this study were run with a fixed Doppler width of 2\,\kms.

The spectral properties at nebular times are complex and sensitive to numerous effects. Most of the low-energy radiation (falling primarily in the optical and near-infrared ranges) in our simulations emerges from the H-rich shell, even when the majority of the decay power is absorbed by the metal-rich layers, located in the deeper layers of the ejecta. Radiation below about 6000\,\AA\ and emitted deep in the ejecta is reprocessed by the H-rich material before it escapes. Although the conditions are nebular, there are still significant optical depth effects at 300\,d after explosion.

The mixing of \nifs\ governs the spatial distribution of the decay power absorbed. Our simulations are particularly sensitive to this distribution because we retain the original progenitor shell stratification, even when mixing is applied. Thus, enhanced \nifs\ mixing favors deposition in the H-rich shell at the expense of the Si-rich material. This effect is a function of progenitor mass since the metal-rich material is less abundant (He-core mass is smaller) in a lower mass massive star. Adopting a complete mixing of the metal-rich layers would probably reduce this effect since this stratification would disappear (this aspect will be examined in a forthcoming study). Furthermore, enhanced deposition in the H-rich envelope systematically leads to an over-ionization of Ca and the weakening of the \caiidoub\ doublet strength (this perhaps takes place in SN\,2012ec; \citealt{jerkstrand_ni_15}). This effect may serve as a means to constrain the level of \nifs\ mixing in the ejecta. Clumping may reduce this over-ionization \citep{d18_fcl}.

Varying the mass of the H-rich shell (as arises from variations in progenitor mass loss or progenitor mass) while keeping the same ejecta kinetic energy changes the chemical stratification in velocity space, the density in the metal-rich core, the absorbed decay power in the H-rich shell, or the trapping efficiency of $\gamma$-rays. This impacts primarily the H\one\ lines in our calculations.

Varying the \nifs\ mass has far reaching consequences since it changes the luminosity, the electron density (and thus the optical depth), and the ionization. The spectral appearance is thus strongly altered. For a high \nifs\ mass, the electron density may become so large that it inhibits the formation of forbidden lines. It is therefore questionable to use SN\,1987A as a template for estimating the yields or the progenitor mass of other SNe II -- there is no linear scaling with \nifs\ nor with the \oidoub\ doublet flux. In the nebular phase, the impact of a varying \nifs\ mass is similar to the impact of time passing for a given SN model.

Although previously reported, for example, by \citet{fransson_chevalier_89}, we reemphasize the influence of the O/Ca abundance ratio in the O-rich shell. Stellar evolution calculations with \mesa\ frequently exhibit the merging of the Si-rich and the O-rich shell during Si burning (see also \citealt{collins_presn_18}), causing the Ca abundance to rise by a factor of about 100 in the O-rich shell. Because \caiidoub\ is a much stronger coolant than \oidoub, this inhibits the production of \oidoub\ at nebular times, destroying any robust relationship between the \oidoub\ doublet flux and the O-rich shell mass (and thus progenitor mass). In Nature, the complete merging of the Si-rich and O-rich shells is probably unlikely, but some contamination of the O-rich shell by Ca-rich material from the Si-rich shell is a possibility (caused by convection and overshoot) and could introduce some variations in the \oidoub\ doublet strength for a given progenitor composition. If the process of shell merging had a higher occurrence rate in higher mass progenitors (owing to the more violent convection that takes place in their interiors), it would provide a natural explanation for the lack of SNe II with strong \oidoub\ emission at nebular times, and inferred progenitor masses below 17\,\msun.

Because we simplified the composition and the associated model atoms, we will have missed some coolants and thus may overestimate the cooling power of some of the lines that we do include. For example, our simulations show a systematic increase in the \oidoub\ doublet luminosity with ZAMS mass, largely irrespective of the adopted mixing, from 2\% (model m12), to 5\% (model m15), 12\% (model m20), and $\sim$\,16\% (model m25) of the bolometric luminosity (or \cofs\ decay power absorbed). This is typically a factor of two greater than obtained by \citet{jerkstrand_ni_15}.

Overall, our simulations suggest that the \oidoub\ doublet strength is the most robust indicator of progenitor mass. The O-rich shell is the most massive metal-rich shell in 12-25\,\msun\ progenitors. Its mass grows considerably with ZAMS mass and thus its associated material captures a large fraction of the decay power. Furthermore,  oxygen is generally neutral under a wide range of ejecta conditions (mixing, \nifs\ mass, progenitor mass), while we obtain a very complicated behavior for Ca (strength dependent on ionization, \nifs\ mixing, line optical depth, etc.). We must however tone down this conclusion since the potential Ca pollution from the Si-rich shell into the O-rich shell can mitigate the \oidoub\ doublet flux at nebular times, making a type II SN appear as if it arose from a lower mass star without such Ca pollution. The implication of this process for progenitor mass estimates is profound and currently understated.

\begin{acknowledgements}

We thank Bill Paxton and Frank Timmes for their help with \mesa, specifically in setting up the proper controls that prevent shell merging during Si burning. Hillier thanks NASA for partial support through the astrophysical theory grant 80NSSC20K0524. We thank the referee, Anders Jerkstrand, for their thoughtful comments on an earlier version
of this paper. This work was granted access to the HPC resources of  CINES under the allocation  2018 -- A0050410554 and 2019 -- A0070410554 made by GENCI, France. This research has made use of NASA's Astrophysics Data System Bibliographic Services.

\end{acknowledgements}

% \bibliographystyle{aa}
% \bibliography{/Users/ldessart/Bibliography/new_sn_library_luc,neb_pap_add}

\end{document}